\documentclass[acmsmall]{acmart}

\AtBeginDocument{%
  \providecommand\BibTeX{{%
    \normalfont B\kern-0.5em{\scshape i\kern-0.25em b}\kern-0.8em\TeX}}}

\usepackage{algorithm}
\usepackage{algorithmicx}
\usepackage{amsmath,amsfonts}
\usepackage{array}
\usepackage{balance}
 \usepackage{booktabs}
\usepackage[skip=1pt,labelfont=bf]{caption}
 \usepackage{calligra}
\usepackage{color}
\usepackage{colortbl}
\usepackage{courier}
\usepackage{csvsimple}
\usepackage{enumitem}
\usepackage{fancybox}
\usepackage{fontenc}
\usepackage{graphicx}
\usepackage{listings}
\usepackage{longtable}
\usepackage{lscape}
\usepackage{makecell}
\usepackage{marvosym}
\usepackage{moreverb}
\usepackage{multicol}
\usepackage{multirow}
\usepackage{pifont}%
\usepackage{rotating}
\usepackage{setspace}
\usepackage{subfigure}
\usepackage[most]{tcolorbox}
\usepackage{threeparttable}
\usepackage{tikz}
\usepackage[normalem]{ulem}
\usepackage{url}
\usepackage{soul}
\usepackage{wasysym}
\usepackage{xspace}

\algnewcommand\algorithmicforeach{\textbf{for each}}
\algdef{S}[FOR]{ForEach}[1]{\algorithmicforeach\ #1\ \algorithmicdo}

\newcolumntype{L}[1]{>{\raggedright\let\newline\\\arraybackslash\hspace{0pt}}m{#1}}
\newcolumntype{C}[1]{>{\centering\let\newline\\\arraybackslash\hspace{0pt}}m{#1}}
\newcolumntype{R}[1]{>{\raggedleft\let\newline\\\arraybackslash\hspace{0pt}}m{#1}}

\definecolor{codegreen}{rgb}{0,0.6,0}
\definecolor{codered}{rgb}{1,0,0}
\definecolor{codegray}{rgb}{0.5,0.5,0.5}
\definecolor{codepurple}{rgb}{0.58,0,0.82}
\definecolor{backcolour}{rgb}{0.95,0.95,0.92}
\definecolor{lightgray}{gray}{0.9}

\lstdefinestyle{mystyle}{
    commentstyle=\color{codegreen},
    keywordstyle=\color{magenta},
    numberstyle=\small\color{black},
    stringstyle=\color{codepurple},
    basicstyle=\scriptsize\ttfamily,
    breakatwhitespace=false,
    breaklines=true,
    captionpos=b,
    keepspaces=true,
    showspaces=false,
    showstringspaces=false,
    showtabs=false,
    tabsize=2
}

\lstdefinelanguage{diff}{
  morecomment=[f][\color{blue}]{@@},     %
  morecomment=[f][\color{red}]-,         %
  morecomment=[f][\color{codegreen}]+,       %
  morecomment=[f][\color{red}]{---}, %
  morecomment=[f][\color{codegreen}]{+++},
}

\setlist{noitemsep} %

\definecolor{darkpastelred}{rgb}{0.76, 0.23, 0.13}
\definecolor{ao(english)}{rgb}{0.0, 0.5, 0.0}

\definecolor{darkpastelred}{rgb}{0.76, 0.23, 0.13}
\definecolor{ao(english)}{rgb}{0.0, 0.5, 0.0}

\definecolor{yellow}{RGB}{255,255,153}
\definecolor{grey}{RGB}{224,224,224}

\newboolean{showcomments}
\setboolean{showcomments}{true}
\ifthenelse{\boolean{showcomments}}
 { \newcommand{\mynote}[2]{
      \fbox{\bfseries\sffamily\scriptsize#1}
        {\small$\blacktriangleright$\textsf{\emph{#2}}$\blacktriangleleft$}}}
        { \newcommand{\mynote}[2]{}}

\definecolor{DarkOrange}{rgb}{0.8,0.3,0.0}
\definecolor{DarkCyan}{rgb}{0.0, 0.55, 0.55}
\definecolor{DarkCyel}{rgb}{1.0, 0.49, 0.0}
\definecolor{yellow-green}{rgb}{0.6, 0.8, 0.2}

\newcolumntype{?}{!{\vrule width 1pt}}

\newcommand{\toolname}{\textsc{BATS}\xspace}

\newcommand{\find}[1]{
\begin{tcolorbox}[leftrule=1mm,toprule=0mm,bottomrule=0mm,left=1pt,right=2pt,top=2pt,bottom=2pt]%
\em #1
\end{tcolorbox}
}

\newcommand{\abox}[1]{
\begin{tcolorbox}[colback=white,boxrule=1pt,top=0pt,bottom=0pt,left=1pt,right=2pt,top=2pt,bottom=2pt]%
\em #1
\end{tcolorbox}
}

\RequirePackage[normalem]{ulem} %
\RequirePackage{color}\definecolor{RED}{rgb}{1,0,0}\definecolor{BLUE}{rgb}{0,0,1} %

\setcopyright{rightsretained}
\acmJournal{TOSEM}
\acmYear{2022} \acmVolume{1} \acmNumber{1} \acmArticle{1}
\acmMonth{1} \acmPrice{}\acmDOI{10.1145/3511096}

\begin{document}

\title{Predicting Patch Correctness Based on the Similarity of Failing Test Cases}

\author{Haoye Tian}
\email{haoye.tian@uni.lu}
\author{Yinghua Li}
\email{yinghua.li@uni.lu}
\author{Weiguo Pian}
\email{weiguo.pian@uni.lu}
\author{Abdoul Kader Kaboré}
\email{abdoulkader.kabore@uni.lu}
\affiliation{%
   \institution{University of Luxembourg}
 	\country{Luxembourg}
}

\author{Kui Liu}\authornote{Corresponding author.}
\email{liukui0811@163.com}
    \affiliation{%
   \institution{Software Engineering Application Technology Lab, Huawei}
 	\country{China}
}

\author{Andrew Habib}
\email{andrew.a.habib@gmail.com}
\author{Jacques Klein}
\email{jacques.klein@uni.lu}
\author{Tegawend\'e F. Bissyand\'e}
\email{tegawende.bissyande@uni.lu}
\affiliation{%
  \institution{University of Luxembourg}
  \country{Luxembourg}
}

\begin{abstract}
 How do we know a generated patch is correct? This is a key challenging question that automated program repair (APR) systems struggle to address given the incompleteness of available test suites. 
 Our intuition is that we can triage correct patches by checking whether each generated patch implements code changes (i.e., behaviour) that are relevant to the bug it addresses. Such a bug is commonly specified by a failing test case. Towards predicting patch correctness in APR, we propose a novel yet simple hypothesis on how the link between the patch behaviour and failing test specifications can be drawn: {\em similar failing test cases should require similar patches}. We then propose \toolname, an unsupervised learning-based approach to predict patch correctness by checking patch {\bf B}ehaviour {\bf A}gainst failing {\bf T}est {\bf S}pecification. \toolname exploits deep representation learning models for code and patches: for a given failing test case, the yielded embedding is used to compute similarity metrics in the search for historical similar test cases to identify the associated applied patches, which are then used as a proxy for assessing the correctness of the APR-generated patches. Experimentally, we first validate our hypothesis by assessing whether ground-truth developer patches cluster together in the same way that their associated failing test cases are clustered. Then, after collecting a large dataset of 1,278 plausible patches (written by developers or generated by 32 APR tools), we use \toolname to predict correct patches: \toolname achieves AUC between 0.557 to 0.718 and recall between 0.562 and 0.854 in identifying correct patches. Our approach outperforms state-of-the-art techniques for identifying correct patches without the need for large labeled patch datasets; as is the case with machine learning-based approaches. While \toolname is constrained by the availability of similar test cases, we show that it can still be complementary to existing approaches: when combined with a recent approach that relies on supervised learning, \toolname improves the overall recall in detecting correct patches. We finally show that \toolname is complementary to the state-of-the-art PATCH-SIM dynamic approach for identifying correct patches generated by APR tools.
\end{abstract}

\settopmatter{printacmref=true}

\ccsdesc[500]{Software and its engineering~Software verification and validation}
\ccsdesc[300]{Software and its engineering~Software defect analysis}
\ccsdesc[100]{Software and its engineering~Software testing and debugging}

\keywords{
Program Repair, Patch Correctness, Test Behavior, Patch Semantics
}

\maketitle

\section{Introduction}
Patch overfitting~\cite{qi2015analysis,smith2015cure} is now acknowledged as one of the major blockers to the adoption of automated program repair (APR) by software practitioners. It refers to the fact that APR-generated patches
often overfit to the repair (incomplete) test suite without necessarily generalizing to other test
cases. In short, overfitting patches do not implement the desired behavior that the program developers would expect. Consequently, when a generated patch is validated as passing all test cases in the test suite, it is referred to as a {\bf\em plausible} patch. Its correctness, indeed, must still be decided manually by developers. Given that existing APR approaches generate a large number of plausible patches, most of which are actually incorrect, there is a need to develop automated approaches that can filter out incorrect patches or that can rank the correct ones higher to alleviate the burden of manual assessment. 

Recently, the literature has proposed various heuristics to predict patch correctness. Csuvik~{\em et~al.}~\cite{csuvik2020utilizing} translate some empirical observations into a simple assumption for ranking valid patches: correct patches apply fewer changes than incorrect ones. Xiong~{\em et~al.}~\cite{xiong2018identifying} build on the hypothesis that test case dynamic execution behaviours are different between correct and incorrect patches. Other researchers propose to focus on learning, with static features of patches, to filter out incorrect patches.  
For example, Ye~{\em et~al.}~\cite{ye2021automated} proposed such a supervised learning-based approach after investing in careful engineering of patch features. In contrast, Tian~{\em et~al.}~\cite{tian2020evaluating} relied on deep representation learning of code changes for yielding patch embeddings that are fed to a supervised learning system.

Overall, existing research in patch correctness prediction has provided promising performance~\cite{wang2020automated}. Nevertheless, they suffer from various caveats. On the one hand, the state of the art dynamic-based approaches require the expensive execution of test cases, which unfavorably impacts the efficiency of the patch assessment process. Besides, such approaches are challenged in practice by the oracle problem: given a test case, we do not always have an accurate specification of what the output should be~\cite{tsimpourlas2020learning}. On the other hand, static-based approaches often require a substantial analysis effort to identify adequate features and properties. In addition, machine learning-based approaches require many labeled patch samples (both correct and overfitting patches). They further exhibit issues of generalization beyond the projects they have been trained on~\cite{wang2020automated}.

In their seminal study on patch plausibility and correctness, Qi {\em et al.}~\cite{qi2015analysis} have presented Kali, a system that performs ``repair'' by only removing or skipping code in programs. Kali generated several patches that pass many weak test suites. Our postulate, however, is that Kali-generated patches should be readily-identifiable as {\em plausible but incorrect} in an APR pipeline: it is unlikely that fixing a program that presents a bug in array iteration would require simply removing whole statements. This calls for research to assess the behaviour of the patch (i.e., what it does) against the nature of the bug (i.e., as expressed by the failing test case specification). 

The idea of checking patch behaviour against failing test specification has not yet been fully exploited in the literature. Recent work by Ye~{\em et~al.}~\cite{ye2021automated} and Tian~{\em et~al.}~\cite{tian2020evaluating} do not even reason about the test cases. The approach by Xiong~{\em et~al.}~\cite{xiong2018identifying} builds on heuristics that consider the similarity of execution behaviours of passing test cases on original and patched programs. Their work, however, does not try to answer the specific question of {\bf \em whether the generated correct patch is actually relevant to the failing test case(s) triggering the repair process}. This is the key hypothesis we introduce and validate:

\abox{\bf 
\begin{center}
	``When different programs fail to pass \ul{similar test cases}, it is likely that these programs require \ul{similar code changes}.''
\end{center}
}

Similarity thus becomes a key challenge: syntactic similarity is not sufficient as it would restrict the search to type-1 or type-2 clones. We have to explore similarity measurements that can capture semantic relationships. Recent work~\cite{alon2019code2vec, hoang2020cc2vec, li2020dlfix, alon2018code2seq, huang2020code} in learning distributed representations of code and code changes have been shown to preserve some semantics (beyond token similarity) and have yielded promising models that were effective on a variety of downstream tasks.%

{\bf This paper.} We build on the aforementioned hypothesis to investigate the possibility of predicting patch correctness by clustering test cases and patches. In this paper, we rely on recent approaches for code representation learning to reason about code and patch similarity. 
\begin{itemize}%
	\item[\ding{182}] {\bf[Heuristic]} We propose a novel heuristic on the relationship between patches and their failing test cases. Although the intuition behind this heuristic is basic and hinted at in regression testing literature~\cite{boehme2014automated}, we are the first to propose and validate it in the APR patch assessment area. %
	
	\item[\ding{183}] {\bf[Validation]} We present a comprehensive validation of the hypothesis that similar test cases are associated with similar patches. Concretely, we consider the case of developer-written patches in the Defects4J dataset and leverag various distance metrics to perform hierarchical clustering based on the embeddings of test cases and patches. 
	
	\item[\ding{184}] {\bf[BATS]} Based on the heuristic, we propose \toolname (Behaviour Against failing Test Specification), an approach to predict patch correctness by statically checking the similarity of generated patches against past correct patches that correspond to failing test cases which are similar to the failing tests of the bug under resolution. More specifically, given one buggy program with its failing test cases and the APR-generated patches, \toolname first enumerates similar failing test cases within the search space of historical bugs. Then, \toolname calculates the similarity between the correct patches associated with the identified failing test cases and the APR-generated patches. Finally, APR-generated patches with similarity scores above a predefined threshold {\em t} are predicted as correct while patches with similarity scores lower than {\em t} are predicted as incorrect. The artifact of this study is publicly available at {\bf\url{https://github.com/HaoyeTianCoder/BATS}}.

	\item[\ding{185}] {\bf[Evaluation]} After collecting a large dataset of plausible patches generated by 32 APR tools or extracted from defects benchmarks, we apply \toolname and measure its performance in identifying correct patches. 
	\begin{itemize}
	    \item Overall, \toolname achieves an AUC (Area Under Curve) between $\sim$0.56 and $\sim$0.72 and a recall between $\sim$56\% $\sim$84\% in identifying correct patches. 
	    \item When comparing with a recent supervised learning classifier~\cite{tian2020evaluating}, \toolname improves the recall in identifying correct patches by 7 percentage points and achieves an equivalent recall in excluding incorrect patches.
	    \item Comparing against the state-of-the-art dynamic approach PATCH-SIM~\cite{xiong2018identifying}, \toolname outperforms it with higher scores of AUC, F1 +Recall and -Recall for the subset of patches where \toolname can find similar test cases.
	\item We note that the performance of \toolname is impacted by the search space for finding similar test cases. Therefore, after demonstrating the promise of the proposed heuristic, we show that it is worthwhile to combine \toolname with the state-of-the-art approaches in order to improve performance in identifying correct patches. To that end we consider a usage scenario where similar test cases are lacking to apply \toolname. Instead, we investigate whether \toolname can still be used as a supplement to another approach: 
	when \toolname is integrated with the recent supervised learning classifier~\cite{tian2020evaluating} , the overall recall in detecting correct patches is improved with 5 percentage points. 
	The experimental results also show that \toolname can be complementary to  PATCH-SIM~\cite{xiong2018identifying} to recall more correct and exclude more incorrect patches.
	\end{itemize}
\end{itemize}

\section{Motivation}
The promise of APR is to automatically find patches for software bugs without human intervention~\cite{monperrus2018automatic,gazzola2019automatic,goues2019automated}. This often translates into a generate-and-validate pipeline~\cite{weimer2009automatically,le2012genprog,le2016enhancing,xiong2017precise,jiang2018shaping,wen2018context,chen2017contract,liu2019tbar, liu2019avatar,ghanbari2019practical} that consists of two main activities: (i) {\em Patch generation} which applies various code transformation and search techniques to produce a set of candidate patches, and (ii) {\em Patch validation} which checks that the candidate patches are relevant to the discovered fault with corresponding test cases.
Identifying correct patches among the automatically validated patches, i.e., checking that the functionality of the patched program satisfies the developer's intention, is usually conducted manually. 
Even recent deep-learning based repair approaches~\cite{lutellier2020coconut,wang2017dynamic,chen2019sequencer,white2019sorting} produce patches that require validation, although new research on involving developers early in the process is arising~\cite{gao2020interactive, bohme2020human}.

In a large body of the literature, patch validation is carried out by executing the patched program against the test suite~\cite{chen2021fast,yi2018correlation}. If a single test case leads to a failure, the patch is discarded as invalid. Rothermel {\em et al.}~\cite{rothermel2001prioritizing} have introduced a test case prioritization strategy in order to early-detect invalid candidate patches. In the same vein, Qi {\em et al.}~\cite{qi2013efficient} empirically highlighted that  test cases that have been effective in revealing invalid patches should be re-run earlier than others on new patch candidates. Nevertheless, these efforts are directed towards identifying valid (i.e., plausibly correct) patches w.r.t. the test suites. Indeed, a test suite being only a weak approximation of the program specification, patch validation with test suites does not provide guarantees on patch correctness. In practice, the correctness of APR-generated patches is checked manually by the developers. Even in the literature, assessment procedures done by researchers check correctness by using heuristics to compare plausible patches against developer oracle patches~\cite{liu2020efficiency}. Some works~\cite{yang2017better} propose to generate more test inputs to strengthen the test inputs and provide more confidence in the generated patches. The oracle problem~\cite{jahangirova2016test} in test generation however makes the execution results of the augmented test suite an unreliable criterion of patch correctness. Shariffdeen {\em et al.} \cite{shariffdeen2021concolic} thus proposed to reduce the space of patch candidates by simultaneously traversing the spaces of test inputs and patches with path exploration.
\section{Approach}
\label{sec:approach}

\begin{figure}[!ht]
\centering
	\includegraphics[width=\linewidth]{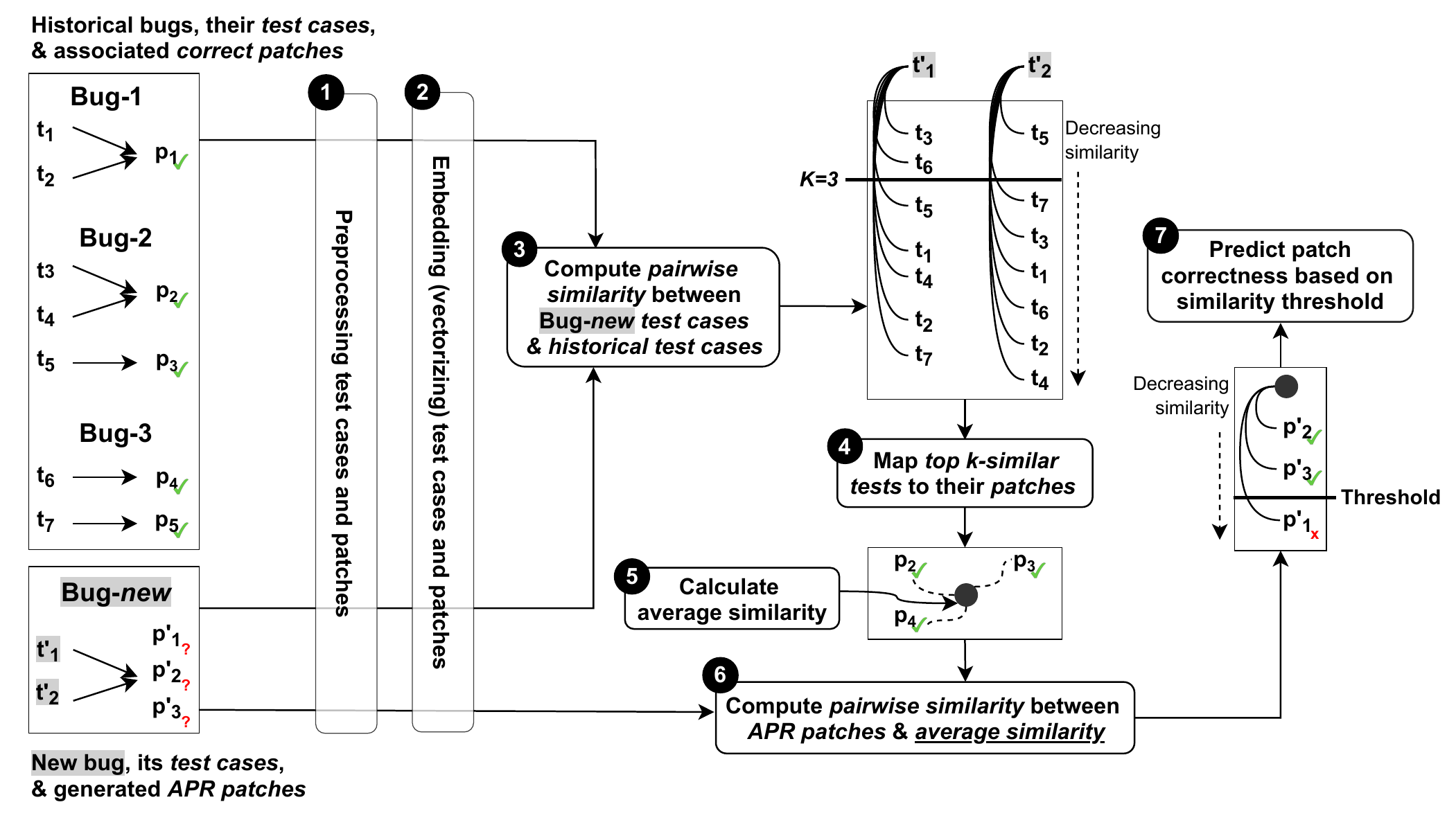}
	\caption{Overview of \toolname.}
	\label{fig:pipeline}
\end{figure} 

In this section, we present \toolname, our approach to predict patch correctness based on the similarity of the new patch to known correct patches with \emph{similar failing test cases}.
Figure~\ref{fig:pipeline} gives an overview of the approach.
\toolname assumes the following inputs:
(i) The bug under resolution, its \emph{associated failing test cases}, and \emph{APR-generated plausible patches}, and
(ii) Historical bugs with \emph{their failing test cases} and the \emph{associated known correct patches}.

To predict patch correctness, \toolname performs the following steps:
(1) Pre-process patches and test cases to prepare them for the embedding step,
(2) Embed patches and test cases into higher dimensional space,
(3) Compute pairwise similarity between \emph{the failing test cases of the bug under resolution} and the \emph{historical test cases},
(4) Select at most top-$k$ historical test cases with similarity score greater than $t_{Test}$, and map them to their associated correct patches,
(5) Compute an average similarity among selected historical known correct patches,
(6) Compute pairwise similarity between every plausible patch for the bug under resolution and the average similarity of known correct patches from the previous step, and
(7) Predict a plausible patch as correct if its similarity to the average similarity of correct patches (from step 5) is greater than some threshold $t_{Patch}$, otherwise, \toolname predicts the patch to be incorrect.

\subsection{Pre-processing Test Cases and Patches}
\toolname leverages the similarity of test cases and their corresponding patches to predict patch correctness.
Therefore, the first two steps of \toolname aim at preparing the test cases and patches into a form suitable for computing similarity.
One way to achieve this is by embedding patches and test cases into higher dimensional space to obtain numerical vectors that are suitable for vector similarity computations.

\paragraph{Tokenizing test cases}
\toolname treats the source code of individual test methods as sequences of tokens while also using camelCase tokenization to further breakdown identifiers into their sub tokens.

\paragraph{Tokenizing patches}
Patches are tokenized in the same manner as test cases with two differences.
First, BATS considers changed lines without their contexts. 
I.e., it selects added and removed lines only, marked with `\textcolor{codegreen}{+}' and `{\textcolor{red}{-}}', respectively.
Second, to keep the information about added and removed lines, \toolname keeps the 
`\textcolor{codegreen}{+}' and `{\textcolor{red}{-}}' markers as part of each patch line.

\subsection{Embedding Test Cases and Patches}
\toolname relies on similarity calculations between test cases and between patches.
Therefore, the second step is to embed test cases and patches into a higher dimensional space to enable similarity computations.
An embedding is a function that maps each token into a high dimension real-value vector while maintaining semantic similarities between similar tokens.

In our approach, embedding individual test methods is straightforward but it is not the case for patches.
Patches are composed of several individual hunks (contiguous changes) while the order of the hunks is irrelevant to the patch.
Each hunk can be embedded as a sequence of tokens into a single vector.
But how can we combine the different hunks to obtain a single vector representing the entire patch?
Simple concatenation of the vectors of different hunks does not produce a unique vector. Because there is no specific order for the individual hunks, different orderings of the different hunks produces different vectors for the same patch.
Therefore, instead of concatenating vectors of different hunks, we sum the vectors of the different hunks to obtain a unique vector representing the entire patch.

To obtain the embeddings for test cases and patches, we leverage three state-of-the-art pre-trained models:
\begin{itemize}[leftmargin=*]
	\item[\ding{119}] {\bf code2vec.} 
     Alon {\em et al.}~\cite{alon2019code2vec} leverage the AST representation of a method to produce its embedding. code2vec has been applied to a variety of downstream tasks, including predicting method names, where it revealed its  ability to learn the structure and semantics of code fragments. We propose in this work to leverage code2vec for embedding test cases. To that end, we build on a pre-trained model provided by the authors~\cite{alon2019code2vec} who trained the model on a dataset containing $\sim$14 million Java methods. 
	
	\item[\ding{119}] {\bf CC2Vec.} 
    Hoang~{\em et~al.}~\cite{hoang2020cc2vec} introduced the CC2Vec hierarchical attention
    neural network model for learning vector representations of patches. In the training phase, the learning is guided by the commit messages that are associated with patches and uses them as semantic descriptions of the patches. 
    We propose in this work to leverage CC2Vec for embedding APR-generated patches. We consider the same architecture where we skip the feature crossing layer and train a new model building on the large dataset of 24,000 patches provided by the authors. 

	\item[\ding{119}] {\bf BERT.} 
	Another popular embedding method that is applied to natural language is BERT~\cite{devlin2019bert}, a transformer-based self-supervised language model. Recent work in software engineering fine-tuned BERT on code fragments and applied it to produce embeddings that were shown effective~\cite{zhou2019lancer,yu2020order}. Tian {\em et al.}~\cite{tian2020evaluating} recently leveraged BERT in their experiments on filtering out incorrect patches based on the similarity between buggy code and patched code. For comprehensive experiments, we also propose to investigate using BERT for embedding patches. %
\end{itemize}

\subsection{Finding Similar Test Cases}
The major hypothesis of \toolname is that similar failing test cases have similar associated patches.
Therefore, the third step of the approach is to find similar test cases from historical fixed bugs, i.e., test cases that failed before their associated fixes are applied, and are similar to the failing test cases of the current bug under resolution.
To this end, \toolname computes pairwise similarities between each failing test of the bug under resolution and all tests found in the search space of \toolname.
To compute the similarity between test cases, we apply the Euclidean distance to their code2vec embeddings.
Then using a similarity threshold, $t_{Test}$, \toolname selects at most the top-$k$ historical tests with similarity score > $t_{Test}$.
If the number of top $k$ similar test cases is smaller than the number of tests with similarity score > $t_{Test}$, $k$ is adjusted accordingly.
Note that historical failing test cases do not need to be from the history of the same project of the bug under resolution.

\subsection{Mapping Historical Failing Test Cases to their Patches}
When \toolname finds the most similar test cases that failed in the past, \toolname further maps these historical failing tests to their associated correct patches which were applied and accepted as correct fixes for those failing tests.
Our hypothesis is that a plausible patch for the current bug under resolution is correct if it is similar to these historical correct patches because their failing test cases are also similar.
To facilitate the comparison of the plausible patches to these historical correct patches\footnote{The patches were written by developers to fix the related bugs and were committed in the historical repository of related projects.}, \toolname averages the historical correct patches by computing an \emph{average} of their embedding vectors.

\subsection{Predicting Patch Correctness}
To predict whether a given plausible patch is correct or not, \toolname calculates the similarity between this patch and the average of the historical correct patches obtained in the previous step.
\toolname computes the similarity of patches through Euclidean and Cosine similarity measurements.
If this similarity score is higher than a threshold $t_{Patch}$, \toolname predicts that this patch is correct, otherwise, the plausible patch is predicted as incorrect.
In our experiments, we set $t_{Patch} = 0.5$.

\subsection{An Example}
Consider the following example of bug Chart-26 from the Defects4J dataset. Chart-26 triggers 22 test cases to fail. To fix this bug, APR tools SOFix and KaliA generate the two plausible patches presented in Figure~\ref{fig:chart26-SOFix} and Figure~\ref{fig:chart26-KaliA}, respectively.

\begin{figure}[!h]
    \centering
    \scriptsize
\begin{lstlisting}[language=Diff,linewidth={\linewidth},frame=tb]
--- .../source/org/jfree/chart/axis/Axis.java
+++ .../source/org/jfree/chart/axis/Axis.java
@@ -1189,10 +1189,12 @@ 
         }
         if (plotState != null && hotspot != null) {
             ChartRenderingInfo owner = plotState.getOwner();
+            if (owner != null) {
                 EntityCollection entities = owner.getEntityCollection();
                 if (entities != null) {
                     entities.add(new AxisLabelEntity(this, hotspot, 
                             this.labelToolTip, this.labelURL));
                 }
+            }
         }
         return state;
\end{lstlisting}
    \caption{A correct patch generated by APR SOFix for the Defects4J bug Chart-26.}
    \label{fig:chart26-SOFix}
\end{figure}

\begin{figure}[!h]
    \centering
    \scriptsize
\begin{lstlisting}[language=Diff,linewidth={\linewidth},frame=tb]
--- .../source/org/jfree/chart/plot/CategoryPlot.java
+++ .../source/org/jfree/chart/plot/CategoryPlot.java
@@ -2541,7 +2541,9 @@
 
         // record the plot area...
         if (state == null) {
         // if the incoming state is null, no information will be passed
+        if (true)
+             return;
         // back to the caller - but we create a temporary state to record
         // the plot area, since that is used later by the axes
         state = new PlotRenderingInfo(null);
\end{lstlisting}
    \caption{An incorrect patch generated by APR KaliA for the Defects4J bug Chart-26.}
    \label{fig:chart26-KaliA}
\end{figure}

First, to find out whether any of the two patches is correct, \toolname looks for test cases that are similar to the 22 failing test cases in the available search space. The search space includes the history of the Chart project as well as other projects, when available. Second, \toolname, thanks to its code2vec-based similarity checker, identifies two test cases as similar to some of the 22 failing tests: one test case is associated with bug Chart-4 and the other is associated with bug Chart-25. A manual investigation of the semantics of these two test cases reveals that they indeed aim at detecting unhandled Null pointer dereferences. 

Third, after \toolname maps the two identified historical test cases to their corresponding correct patches, it measures the similarity of the APR-generated patches to these relevant correct patches. Finally, based on the similarity score and a similarity threshold, \toolname precisely predicts that the generated patches by SOFix and KaliA are correct and incorrect, respectively. When we manually inspect the historical correct patches that were applied to fix Chart-4 (cf. Figure~\ref{fig:chart4developer}) and Chart-25 (cf. Figure~\ref{fig:chart25developer}), we notice that they both implement similar behavior (i.e., adding null check) as the proposed patch by SOFix (cf. Figure\ref{fig:chart26-SOFix}). On the contrary, the KaliA patch (cf. Figure~\ref{fig:chart26-KaliA}) suggests an irrelevant code change.

\begin{figure}[!t]
    \centering
    \scriptsize
\begin{lstlisting}[language=Diff,linewidth={\linewidth},frame=tb]
--- .../source/org/jfree/chart/plot/XYPlot.java
+++ .../source/org/jfree/chart/plot/XYPlot.java
@@ -4490,6 +4490,7 @@ public class XYPlot extends Plot implements ValueAxisPlot, Pannable,
                     }
                 }
                 
+                if (r != null) {
                     Collection c = r.getAnnotations();
                     Iterator i = c.iterator();
                     while (i.hasNext()) {
@@ -4498,6 +4499,7 @@ public class XYPlot extends Plot implements ValueAxisPlot, Pannable,
                             includedAnnotations.add(a);
                         }
                     }
+                }
             }
         }
\end{lstlisting}
    \caption{A correct developer-written patch for the Defects4J bug Chart-4.}
    \label{fig:chart4developer}
\end{figure}
\begin{figure}[!t]
    \centering
    \scriptsize
\begin{lstlisting}[language=Diff,linewidth={\linewidth},frame=tb]
--- .../source/org/jfree/chart/renderer/category/StatisticalBarRenderer.java
+++ .../source/org/jfree/chart/renderer/category/StatisticalBarRenderer.java
@@ -256,6 +256,9 @@ public class StatisticalBarRenderer extends BarRenderer
 
         // BAR X
         Number meanValue = dataset.getMeanValue(row, column);
+        if (meanValue == null) {
+            return;
+        }
 
         double value = meanValue.doubleValue();
         double base = 0.0;
@@ -312,7 +315,9 @@ public class StatisticalBarRenderer extends BarRenderer
         }
 
         // standard deviation lines
-            double valueDelta = dataset.getStdDevValue(row, column).doubleValue();
+        Number n = dataset.getStdDevValue(row, column);
+        if (n != null) {
+            double valueDelta = n.doubleValue();
             double highVal = rangeAxis.valueToJava2D(meanValue.doubleValue() 
                     + valueDelta, dataArea, yAxisLocation);
             double lowVal = rangeAxis.valueToJava2D(meanValue.doubleValue() 
@@ -341,6 +346,7 @@ public class StatisticalBarRenderer extends BarRenderer
             line = new Line2D.Double(lowVal, rectY + rectHeight * 0.25, 
                                      lowVal, rectY + rectHeight * 0.75);
             g2.draw(line);
+        }
         
         CategoryItemLabelGenerator generator = getItemLabelGenerator(row, 
                 column);
\end{lstlisting}
    \caption{A correct developer-written patch for the Defects4J bug Chart-25.}
    \label{fig:chart25developer}
\end{figure}

\begin{figure}[!ht]
\setlength{\abovecaptionskip}{0.5mm} 
\centering
	\includegraphics[width=1\linewidth]{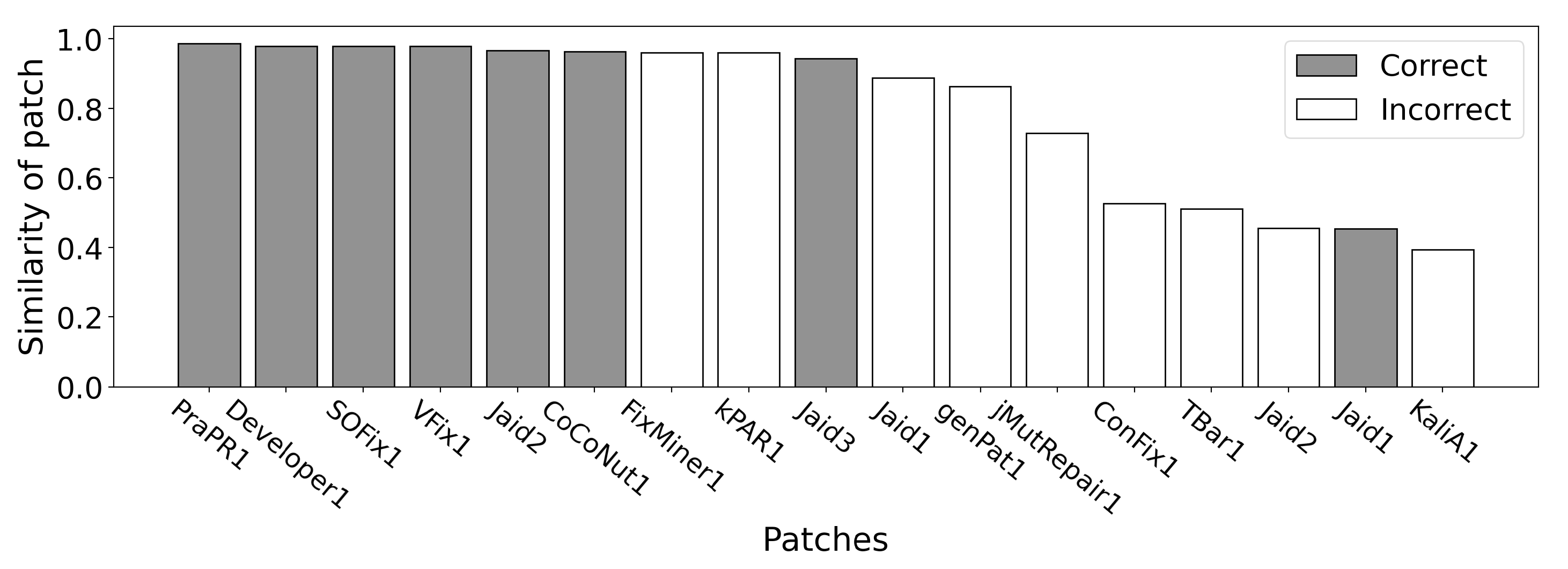}
	\caption{The ranked patches generated by APR tools for Chart-26. The numerical value next to each tool name indicates patch id since a tool can generate more than one patch.}
	\label{fig:rank}
\end{figure}

In our evaluation, there are 17 plausible patches generated by different APR tools for Chart-26\footnote{These APR-generated patches have been labeled in previous work.}. In Figure~\ref{fig:rank}, we see the 17 patches ranked according to their similarity score computed by \toolname. On the x-axis, each patch is labeled with a combination of the name of the APR tool that generated it and a numerical id as a tool may generate more than one patch. We note that most of the correct patches (grey bars) are ranked ahead of incorrect patches (white bars) which confirms that our hypothesis is effective in discriminating correct patches from incorrect ones.

\section{Experimental Setup}
In this section, we introduce the experimental setup to evaluate our hypothesis and our approach, \toolname. We present the research questions in Section~\ref{sec:rqs}, the datasets in Section~\ref{sec:datasets}, and the evaluation metrics in Sections~\ref{section:metrics} and~\ref{sec:metrics}.

\subsection{Research Questions}
\label{sec:rqs}
Our research questions aim to validate the hypothesis, draw insights for developing a prediction method for patch correctness and finally assess the \toolname approach with APR-generated plausible patches while comparing against recent state of the art approaches.
\begin{description}
    \item {\em {\bf RQ-1.} Does the similarity of bug-triggering test cases correlate with the similarity of the associated bug fixing patches?} This research question aims to validate the feasibility of our proposed hypothesis. To this end, we conduct two experiments: clustering similar test cases and assessing patch similarity with the similar test cases. This offers insights into the design of the patch correctness identification system based on inferred thresholds.
	\item {\em {\bf RQ-2.} To what extent can an approach assessing patch behaviour against test specification based on unsupervised learning be effective in identifying correct patches among plausible ones?} With this research question, we first present the implementation of \toolname and evaluate its performance in identifying correct patches in 1,278 patches generated by 32 APR tools. 
	\item {\bf RQ-3.} {\em Can \toolname achieve competitive results against the recent state of the art approaches?} In this research question, we compare \toolname against static and dynamic approaches of predicting patch correctness for APR tools. Then we explore the possibility of leveraging \toolname to complement the state of the art in identifying correct patches.
\end{description}

\subsection{Datasets}
\label{sec:datasets}

We focus our experiments on the Defects4J~\cite{just2014defects4j} benchmark since it is widely used in the literature, and we can readily collect plausible patches generated by APR tools on the programs included in the benchmark. Table~\ref{tab:patches} provides the statistics on the collected patches. Besides the 205 developer (correct) patches provided in the benchmark, we also leverage the reproduced dataset from the study of 16 APR systems by Liu et al.~\cite{liu2020efficiency}, which we augment with a dataset provided by Ye et al.~\cite{ye2021automatedPatch}. Finally, we also scan the artifacts released in the literature towards identifying plausible patches generated by recent APR tools. Overall, we share with the community the largest dataset\footnote{\url{https://github.com/HaoyeTianCoder/BATS}}, to-date, of patches for Defects4J bugs, which includes hundreds of overfitting (i.e., incorrect) patches.

\begin{table}[!t]
	\centering
	\caption{Statistics on the dataset of developer-written and APR-generated patches.}
	\label{tab:patches}
	\resizebox{1\linewidth}{!}
	{
	\begin{threeparttable}

		\begin{tabular}{lrrr|lrrr}
			\toprule
            {\bf Subject} & {\bf Patches} & {\bf Incorrect}  &{\bf Correct} & {\bf Subject} & {\bf Patches} & {\bf Incorrect}  &{\bf Correct} \\\midrule
            {\bf Developer}~\cite{just2014defects4j} & 205 & 0  & 205 & {\bf jGenProg2015}~\cite{durieux2015automatic}& 11& 8  & 3   \\
            {\bf 3sFix}~\cite{chen2018remarkable} & 60  & 57 & 3   & {\bf jKali}~\cite{martinez2016astor} & 14  & 12 & 2   \\
            {\bf ACS}~\cite{xiong2017precise} & 21  & 6  & 15  & {\bf jMutRepair}~\cite{martinez2016astor} & 15  & 12 & 3   \\
            {\bf ARJA}~\cite{yuan2020arja} & 183 &168 & 15  & {\bf KaliA}~\cite{yuan2020arja} & 14  & 14 & 0   \\
            {\bf CapGen}~\cite{wen2018context} & 64  & 39 & 25  & {\bf kPAR}~\cite{liu2019you} & 50  & 42 & 8   \\
            {\bf Cardumen}~\cite{martinez2018ultra} & 10  & 8  & 2   & {\bf LSRepair}~\cite{liu2018lsrepair} & 18  & 15 & 3   \\
            {\bf CoCoNut}~\cite{lutellier2020coconut} & 28  & 0  & 28  & {\bf Nopol2015}~\cite{durieux2015automatic} & 10  &  8 & 2   \\
            {\bf ConFix}~\cite{kim2019automatic} & 66  & 52 & 14  & {\bf PraPR}~\cite{ghanbari2019practical} & 22  & 0  & 22  \\
            {\bf DeepRepair}~\cite{white2019sorting}& 13  & 9  & 4   & {\bf RSRepiarA}~\cite{yuan2020arja} & 18  & 18 & 0   \\
            {\bf DynaMoth}~\cite{durieux2016dynamoth} & 18  & 18 & 0   & {\bf SequenceR} ~\cite{chen2019sequencer} & 35  & 24 & 11  \\
            {\bf ELIXIR}~\cite{saha2017elixir} & 36  & 13 & 23  & {\bf SimFix}~\cite{jiang2018shaping} & 35  & 12 & 23  \\
            {\bf FixMiner}~\cite{koyuncu2020fixminer} & 27  & 17 & 10  & {\bf SketchFix}~\cite{hua2018towards} & 21  & 9  & 12  \\
            {\bf GenPat}~\cite{jiang2019inferring} & 29  & 18 & 11  & {\bf SOFix}~\cite{liu2018mining} & 21 & 2  & 19  \\
            {\bf GenProgA}~\cite{yuan2020arja} & 14  & 14 & 0   & {\bf ssFix}~\cite{xin2017leveraging} & 19  & 8  & 11  \\
            {\bf HDRepair}~\cite{le2016history} & 6   & 2  & 4   & {\bf TBar}~\cite{liu2019tbar} & 57  & 36 & 21  \\
            {\bf Hercules}~\cite{saha2019harnessing} & 51  & 14 & 37  & {\bf VFix}~\cite{xu2019vfix} & 23  &  1 & 22  \\
            {\bf JAID}~\cite{chen2017contract} & 64  & 33 & 31  & & & & \\\midrule%
            & & & & {\bf All}      & 1,278 & 689 & 589 \\            
			\bottomrule
		\end{tabular}
	\end{threeparttable}
	}
\end{table}

{\bf Dataset per experiment:} 
For answering RQ-1, we rely on the 1,120 failing test cases and the associated 205 developer patches in the Defects4J dataset. For answering RQ-2 and RQ-3 (assessment of the \toolname approach), we consider the 1,278 plausible patches generated by APR tools or provided by the Defects4j project developers. In these RQs, we also rely on the 1,120 test cases and 205 developer patches as the search space for similar test cases to the failing test case being addressed. We ensure, however, that for every execution of \toolname we remove from the search space the failing test cases and the developer patches that are related to the bug under resolution.

\begin{figure}[!t]
\centering
	\includegraphics[width=0.75\linewidth,]{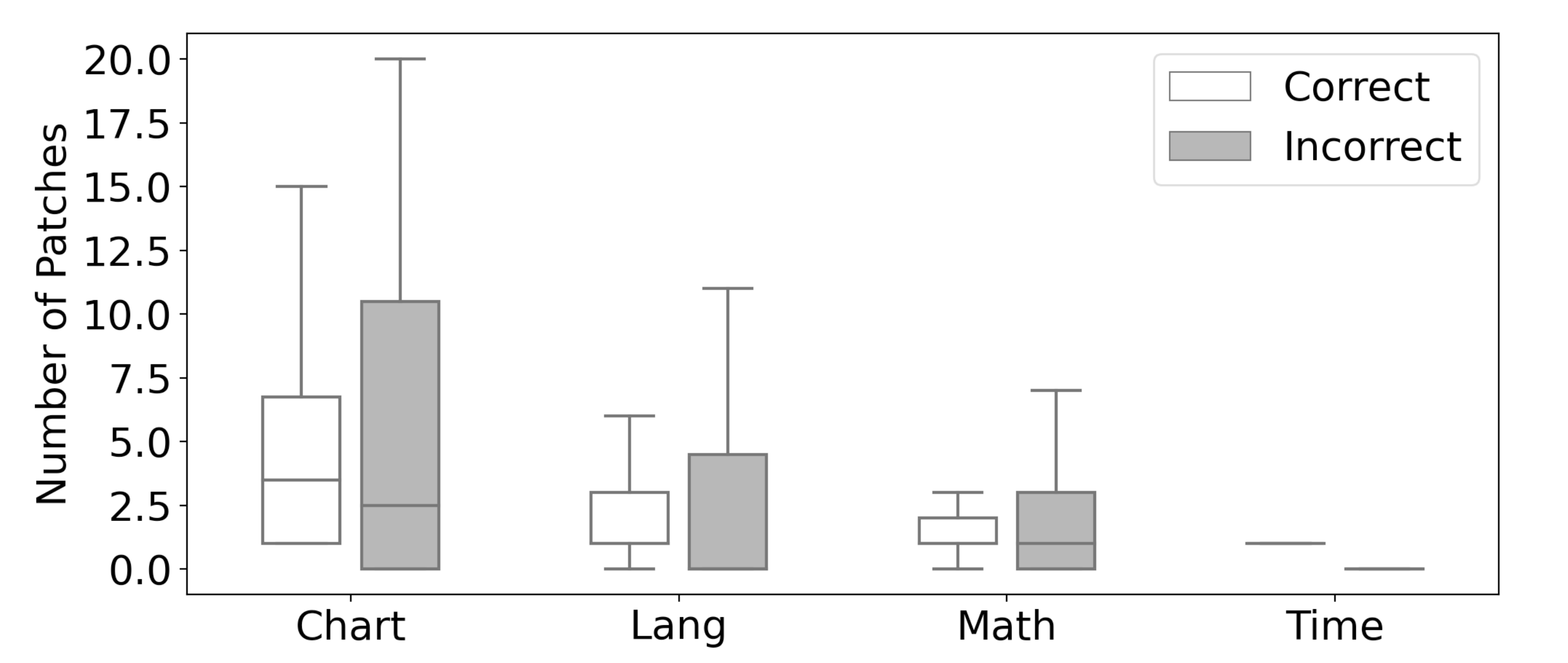}
	\caption{Distribution of the number of collected patches per project in the Defects4j dataset.}
	\label{fig:boxplot_patch}
\end{figure}

Figure~\ref{fig:boxplot_patch} presents the distribution of 1,278 generated patches for Chart, Lang, Math, and Time projects that  have been widely used in the community of automated program repair~\cite{liu2020efficiency,liu2021critical} and patch correctness identification~\cite{xiong2018identifying,tian2020evaluating}.

\subsection{Cluster Analysis Metrics}
\label{section:metrics}
To group similar test cases and similar patches together respectively, we explore unsupervised learning algorithms to perform clustering. K-means~\cite{arthur2006k} generally provides a good clustering performance and strong interpretability. Its main limitation, however, is that it requires the user to specify the number of clusters $k$, which is unfortunately specific to the datasets. We therefore adopt a hierarchical clustering algorithm, e.g., bisecting K-means~\cite{karypis2000comparison}, to empirically determine the appropriate value of $k$.

{\bf Bisecting K-means.} This algorithm was initially proposed to overcome the challenges of K-means (local minima, non-spherical clusters, etc.). Bisecting K-means modifies the K-Means algorithm to produce partitional/hierarchical clustering, thus recognizes clusters of any shape and size. The algorithm starts by splitting the dataset into two clusters based on K-means. Then, iteratively, each cluster is split. Each time the two clusters are identified as those presenting the smaller sum of
squared errors (SSE). By placing a threshold for the performance in terms of the sum of squared errors, the clustering iterations can be stopped. This algorithm is therefore convenient to infer a reasonable number $k$ of clusters in a dataset.

{\bf Cluster Similarity Coefficient (Adjusted Silhouette Coefficient).} Once clusters are yielded, we can measure to what extent each element is actually similar to its own cluster (cohesion) compared to other clusters (separation). To that end, we propose a similarity coefficient (SC) and cluster similarity coefficient (CSC) metrics of each cluster element based on its internal similarity and its external similarity towards other elements. We use the following equations:
\vspace{-6mm}
\begin{multicols}{2}
    \begin{equation}\label{si}
    	SC(e)=\frac{in(e)-out(e)}{\max \{in(e), out(e)\}}
	\end{equation}

	\begin{equation}\label{sc}
    	CSC=\frac{1}{n} \sum_{e=1}^{n} SC(e)
	\end{equation}
\end{multicols}

\noindent
where $in(e)$ represents the average Euclidean Similarity from the (test case or patch) embedding $e$ to other embeddings in the same cluster;  $out(e)$ represents the average Euclidean Similarity from the embedding $e$ to the embeddings in  other clusters. The value range of $SC$ is [-1,1]: the larger the value of $SC$, the better the clustering effect. When $SC$ value is greater than 0, the clustering is consistent. And in order to measure the overall 
performance, $CSC$ (averaging $SC$) is used to calculate the coefficient taking into account all clusters.

{\bf Sum of Squared Error (SSE).}
\label{sec:sse} Finally, we rely on the commonly-used sum of 
squared errors to measure the variance within clusters. SSE in our study is computed as the sum of the squared differences between each embedding and its cluster's mean. 
If  within each  and every cluster, all cases are identical, the SSE would then be equal to 0 as per equation~\ref{sse}.
\begin{equation}
\label{sse}
S S E=\sum_{i=1}^{k} \sum_{j=1}^{n_{i}}\left(x_{i j}-x_{i}\right)^{2}
\end{equation}
where $k$ represents the number of clusters, $n_{i}$ represents the number of elements of the i-th cluster, $x_{ij}$ represents the j-th element of the i-th cluster, and $x_{i}$ represents the center element of the i-th cluster. Therefore, the smaller the SSE, the better the clustering effect.

\subsection{Performance Metrics}
\label{sec:metrics}

We consider the {\bf Recall} of \toolname in two dimensions: 
\begin{itemize}
	\item {\bf +Recall} measures to what extent correct patches are identified, i.e., the percentage of correct patches that are indeed predicted as correct~\cite{tian2020evaluating}.
	\item {\bf -Recall} measures to what extent incorrect patches are filtered out, i.e., the percentage of incorrect patches that are indeed predicted as incorrect~\cite{tian2020evaluating}. 
\end{itemize}

\vspace{-6mm}
\begin{multicols}{2}
    \begin{equation}\label{Recall_P}
    + Recall=\frac{TP}{TP+FN}
    \end{equation}
    
    \begin{equation}\label{Recall_N}
    - Recall=\frac{TN}{TN+FP}
    \end{equation}
\end{multicols}

\noindent
where $TP$ represents true positive, $FN$ represents false negative, $FP$ represents false positive, $TN$ represents true negative.

{\bf Area Under Curve (AUC) and F1}. By considering the similarity score as a prediction probability, \toolname can be evaluated like any machine learning model with the common metrics such as AUC and F1 score (harmonic mean between precision and recall for identifying correct patches).

{\bf MAP and  MRR}. Since \toolname  ranks all generated patches based on similarity scores, instead of simply considering the top-1 as the correct, we can consider a recommendation and leverage common metrics used in assessing the ranked list. The mean average precision (MAP) and mean reciprocal rank (MRR) are such metrics that help assess whether \toolname can place correct patches ahead of incorrect patches in the ranked list presented to the developers. 
\vspace{-6mm}
\begin{multicols}{2}
    \begin{equation}
    M A P=\frac{1}{n} \sum_{i=1}^{n} \frac{\sum_{j=1}^{m}(P_{ij} \times R e l_{ij})}{\text{\# correct patches}}
    \end{equation}
    
    \begin{equation}\label{MRR}
    M R R=\frac{1}{n} \sum_{i=1}^{n} \frac{1}{{rank}_{i}}
    \end{equation}
\end{multicols}
    
\noindent
Where $Rel_{ij}$ is an indicator function that is 1 if the $j$-th patch is correct in list $i$, otherwise it is 0. $P_{ij}$ is the precision at the threshold $j$ in the list $i$, and the $rank$ represents the first correct ranking.

\toolname relies on the unsupervised learning technique and considers to set thresholds of similarities among test cases and patches to predict the correctness of patches.
Given that in practice we cannot tune the decision threshold based on the specific case of each bug, we fix a single similarity threshold to compute the accuracy, precision, and false positives. We set the model prediction threshold to 0.5. As soon as the normalized similarity score between the generated patch and the identified cluster of historical patches is higher than 0.5, we predict the patch as a correct one. Otherwise, we predict it as incorrect. We note, however, that in the literature, some approaches~\cite{xiong2018identifying} are assessed by adjusting the threshold in the test data to achieve the best possible +Recall.

\section{Experimental results}
\label{sec:experiments}
In this section, we first present the validation of the hypothesis in our work (Answer to RQ-1 in Section~\ref{subsec:rq1}). Then %
, we report the experimental results of patch correctness prediction on our collected dataset (Answer to RQ-2 in Section~\ref{subsec:rq2} and RQ-3 in Section~\ref{subsec:rq3}).

\subsection{[RQ-1] Cluster of Similar Test Cases and Patches}
\label{subsec:rq1}

\noindent
{\bf [Objective]:}
We perform experiments to answer RQ-1, whether the proposed hypothesis is valid for patch correctness identification with the following two sub questions to observe whether failing test cases are similar when the associated patches are similar. First, we investigate whether the 1,120 failing test cases in Defects4J can be grouped in clearly separable clusters. Based on each such test cluster, we decide that the associated 205 developer patches constitute a cluster. 
Then, we seek to validate the feasibility of leveraging similar test cases to predict the patch correctness with Defects4J dataset.
\begin{itemize}
	\item {\bf RQ-1.1} {\em Do patches cluster well together when their test cases are similar?} To answer this RQ, we automatically cluster test cases into groups of similar test cases, then we assess whether the associated patches in each group also have a good clustering cohesion.
	\item {\bf RQ-1.2} {\em Given two test cases, can their similarity score be used as an indicator of their relatedness in terms of patch similarity?} We investigate test case similarity vs. patch similarity hypothesis in a fine-grained manner beyond the clusters. %
	\end{itemize}

\noindent
{\bf [Experimental Design for RQ-1.1]:} First, we use the code2vec and CC2Vec pre-trained models to produce embeddings for each test case and for each patch respectively in the Defects4J dataset. To cluster test cases, we rely on the bisecting K-means algorithm to produce hierarchical clusters. 
At each iteration of bisecting K-means, we compute the sum of squared error of the clusters. We use the evolution of reduction in SSE values to decide on a threshold for the number of clusters into which we can split the test cases in our dataset. Experimentally, we observed that the number of clusters $k$ for which the SSE saturates, i.e., no longer drops, is $40$. 
Additionally, we empirically validate the consistency of our proposed hypothesis by investigating the Cluster Similarity Coefficient (CSC) of different cluster settings (i.e., $k \in \{30, 40, 50\}$).

In this experiment, we leverage the Similarity Coefficient (SC) and Cluster Similarity Coefficient (CSC) (cf. Section~\ref{section:metrics}) to assess the consistency and cohesion of the yielded clusters. Nevertheless, we can observe the clustering effect for test cases by investigating the distance distribution of each test case to the center of its cluster: we first consider the distance\footnote{Distance and similarity are two concepts that are used interchangeably in this section depending on the context to facilitate comprehension.} with all test cases in a single group. Then, we compute the distance of each test case to the center of its K-means-inferred cluster. We consider that if the distribution of distances shows that, on average, distances in a cluster are lower than in the whole group, then the test cases have been well-separated. 
We confirm that when considering the whole dataset, the distances are the longest (i.e., the median value of distance distributions for all clusters are lower than the one for the whole dataset). In several inferred clusters, the median distance (of the test case to the cluster center) is halved. These observations suggest indeed that the test cases could be readily clustered.

\noindent
{\bf [Experimental Results for RQ-1.1]:} Considering the test cases clusters, we infer associated groupings of patches that address these test cases. We then compute the CSC to evaluate the consistency and cohesion of the patch clusters that we have derived. These metrics evaluate the distances among patches within each cluster and the distances among clusters. We validate the metrics on the overall Defects4J ground-truth dataset.

Table~\ref{tab:silhouette_multi} presents the cohesion and consistency metrics for the patches regrouped based on the clusters of test cases when the number of clusters is 30, 40 and 50, respectively. The Cluster Similarity Coefficient(CSC) is the average value of similarity coefficient (SC) values for all clusters. ``{\bf Qualified}'' represents the ratio of clusters that have $SC>0$ out of all clusters identified.
We compare those metrics (on test case clusters) to the associated patch clusters. The positive values of the CSC indicate that, on average, the elements inside the same group are indeed more similar among themselves than they are similar to the elements in other groups. 
When test cases are well grouped ($CSC > 0$), the corresponding clustering of the associated patches clusters together  consistently ($CSC > 0$). In more details, a large ratio of clusters (33/40) also have high cohesion. 
When adjusting the number of clusters to 30 or 50, the results show that the major clusters of patches (23/30, 37/50) still keep high cohesion, and the clustering of test cases and patches are consistent to each other ($CSC > 0$).

\begin{table}[!h]
	\centering
	\caption{Statistics on the performance of clustering of test cases and patches with 30, 40 and 50 clusters.}
	\label{tab:silhouette_multi}
	{
	\begin{threeparttable}
		\begin{tabular}{l|r|r}
			\toprule
			{\bf Subjects} & {\bf Cluster Similarity Coefficient} & {\bf Qualified} \\
			\midrule
			\multirow{1}{*}Test Cases & 0.19 & 30/30  \\
			 Patches & 0.16 & 23/30  \\
			 \hline
			\multirow{1}{*}Test Cases & 0.19 & 40/40  \\
			 Patches & 0.16 & 33/40  \\
			 \hline
			 \multirow{1}{*}Test Cases & 0.21 & 50/50  \\
			 Patches & 0.14 & 37/50  \\
			\bottomrule
		\end{tabular}
	\end{threeparttable}
}
\end{table}

\begin{figure}[!t]
\setlength{\abovecaptionskip}{1mm} 
\centering
	\includegraphics[width=1\linewidth]{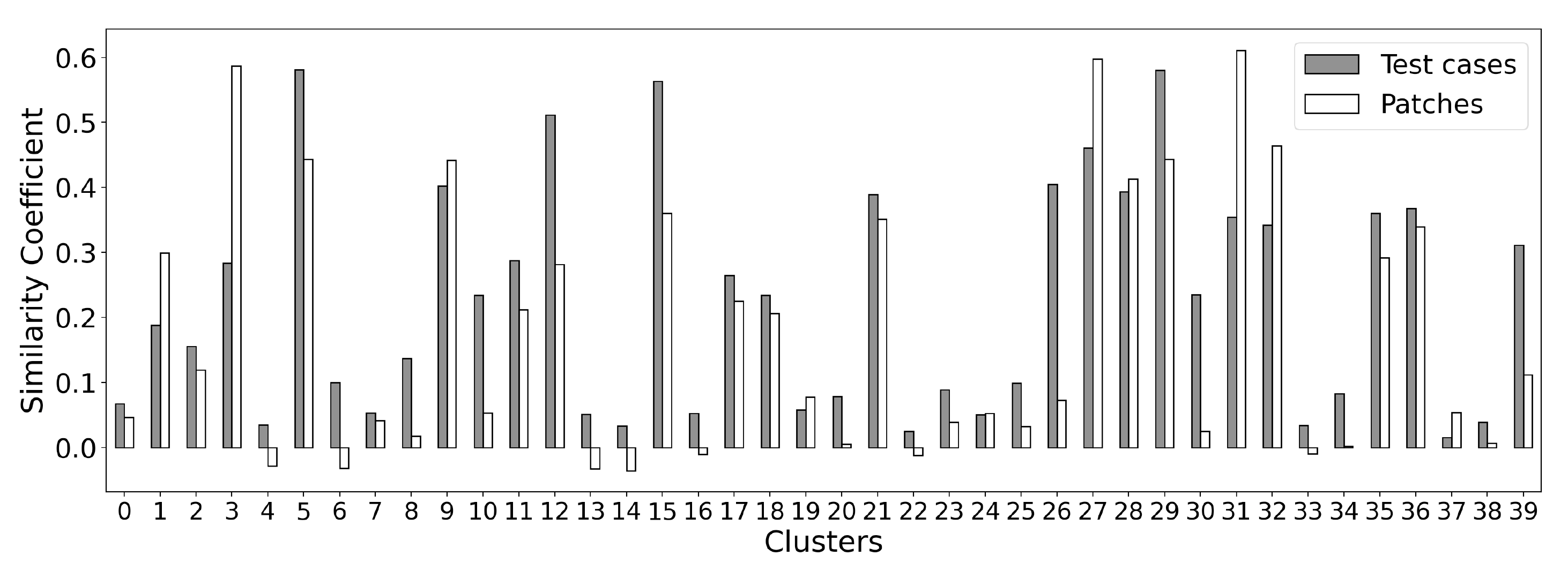}
	\caption{Similarity coefficient of test cases and patches at each cluster.}
	\label{fig:sc_plot}
\end{figure}

Figure~\ref{fig:sc_plot} further presents the similarity coefficient (SC) of test cases and patches for each cluster when $k$ is set to 40. 
We observe that, for most pairwise clusters of test cases and patches, when the SC value of test cases (presented with grey bar) in one cluster is high, the associated patches (presented with white bar) also have a high SC score. 
We further calculate a Pearson correlation between the clusters of test cases and the clusters of associated patches, of which value is 0.883 $>0$.
Pearson correlation can be used to measure the linear correlation between two sets of data where a higher positive value (i.e., $>0$) indicates a more positive association.
Such results indicate that the similar test cases can lead to similar patches.

\find{{\bf [RQ-1.1] } Given a cluster of similar test cases, their associated patches cluster consistently. This experiment also hints that the representation models for test cases and clusters yield meaningful embeddings for investigating the relatedness of patches and test cases.}

\noindent
{\bf [Experimental Design for RQ-1.2]:}
The clustering experiment for RQ-1.1 focuses on average distances within clusters, we further seek to validate the possibility of using test case similarity as a potential heuristic to predict correct patch behavior. The objective is to answer ``{\em whether the similarity of two test cases can be used as an indicator of their relatedness in terms of patch similarity}''.
To this end, we first consider finding the most similar test case from the search space of historical test cases for the failing-executed test case of a given bug.
We then assess to what extent the patch of a given bug is similar to the patch associated to the most similar test case (referred to \ding{172} {\bf Scenario~H}), of which results are compared against the results in \ding{173} {\bf Scenario~N} where we compute the average similarity between a given patch and all other patches.
The experiments finally investigate scenarios where the closest test case is sought within the all project or only in other projects (excluding the one where the test case is found).

\begin{figure}[!t]
\centering
	\includegraphics[width=0.8\linewidth]{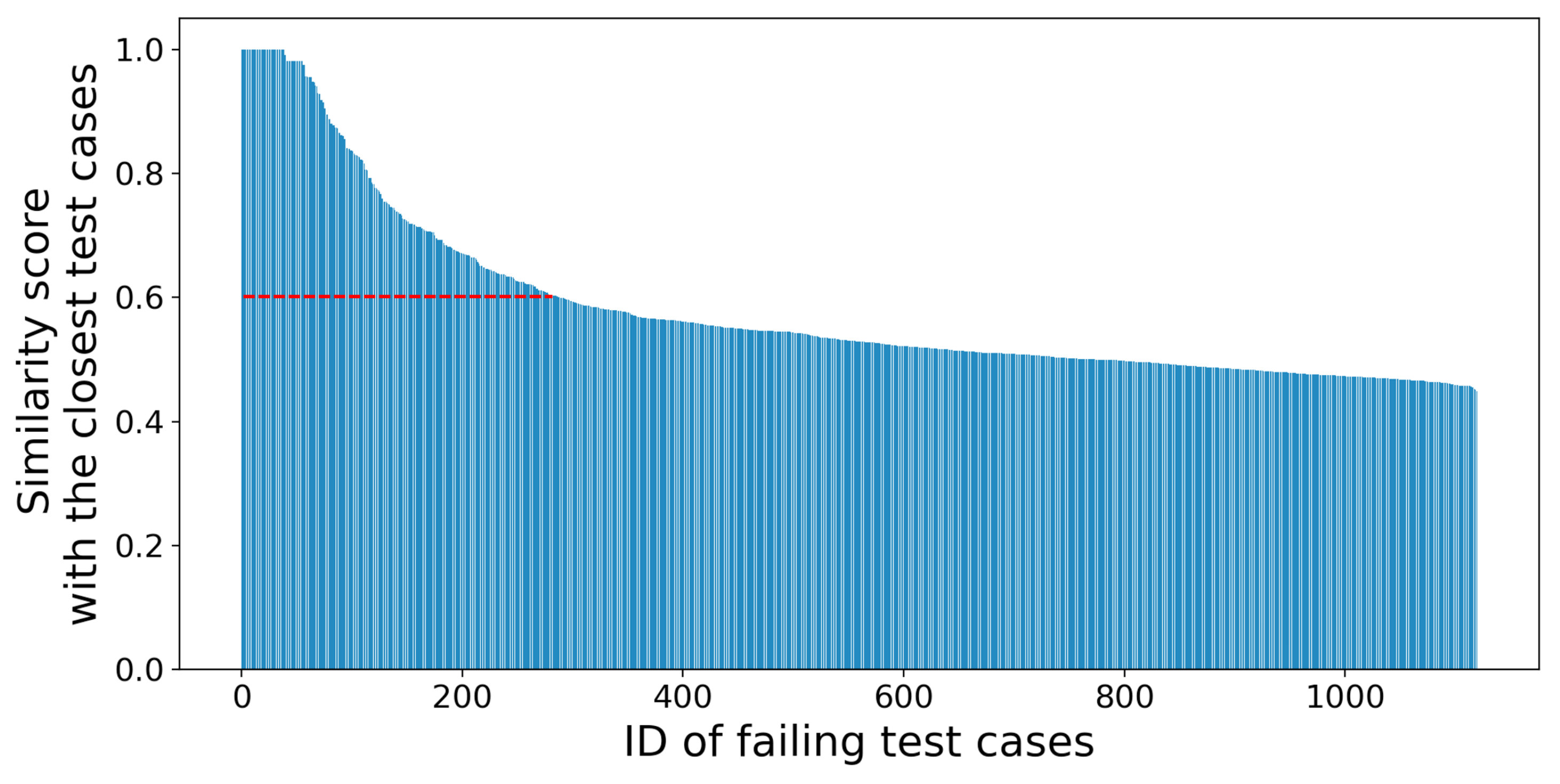}
	\caption{Distribution on the similarities between each failing test case of each bug and its closest similar test case.}
	\label{fig:Similarity_Test}
\end{figure}

\noindent
{\bf [Experimental Results for RQ-1.2]:} 
Figure~\ref{fig:Similarity_Test} presents the overall similarity distribution between each of the 1,120 test cases in the dataset and its closest counterpart: while some test cases indeed have very similar counterparts, many test cases have low similarities with their closest counterparts. 
These relatively low similarities for many test cases can be explained by the limited number of test cases considered in the study datasets. 
This results suggest that it may not always make sense, for a given test case, to blindly consider the most similar counterpart since this counterpart can still be highly dissimilar. 
We thus propose to experimentally determine a threshold to decide when in practice the closest test should not be considered as similar. 

\begin{figure}[!t]
\centering
	\includegraphics[width=0.8\linewidth,]{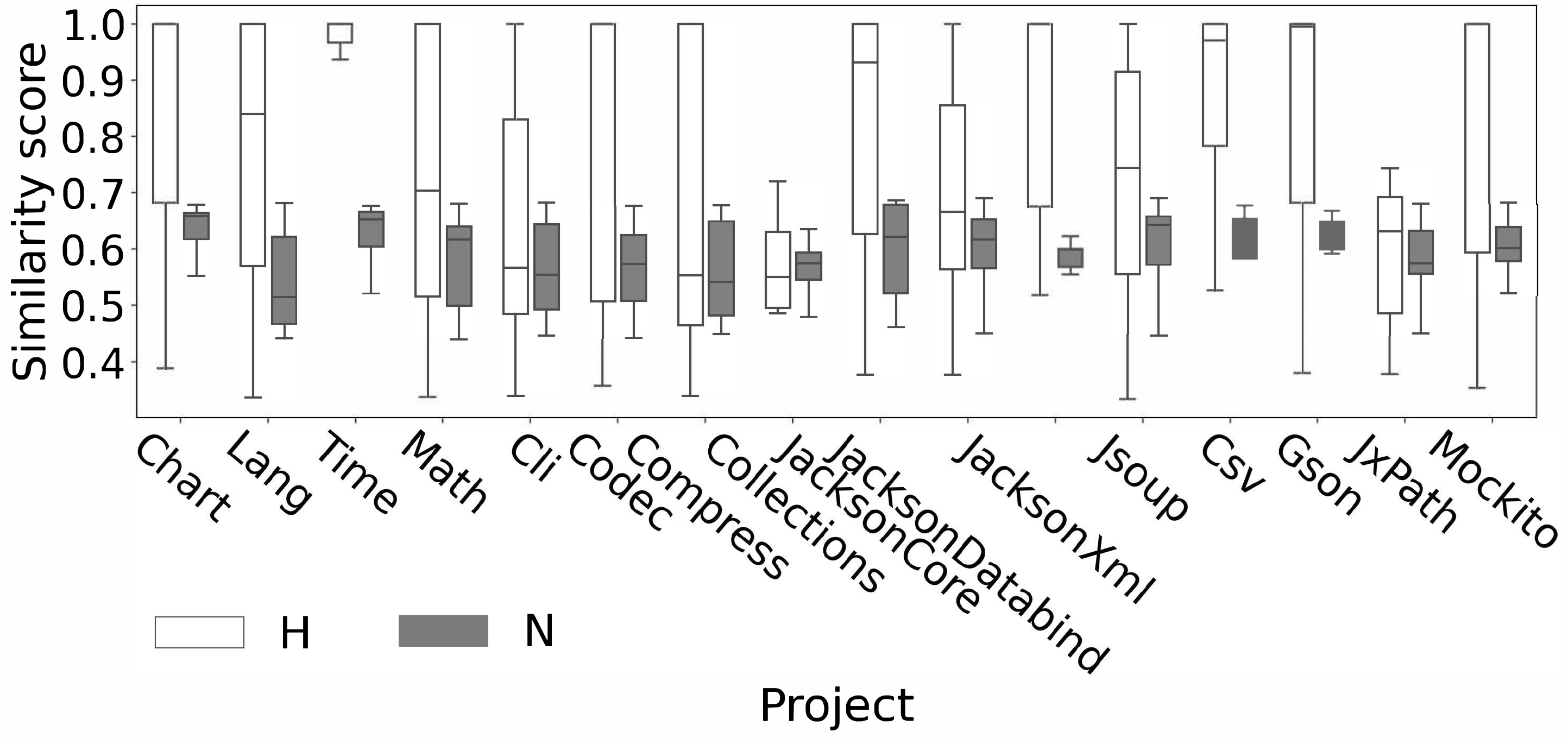}
	\caption{Distributions on the similarities of pairwise patches (similar patch selected with Scenario H vs. Scenario N from \underline{\bf all projects}, i.e., the search space for searching similar cases is all projects in the dataset).}
	\label{fig:boxplot_nofilter}
\end{figure} 

In Figure~\ref{fig:boxplot_nofilter}, we compare the distributions of the similarity scores for the two scenarios H and N. 
In all projects, the distributions in the scenario H of our hypothesis present higher similarities. This indicates that the most similar test case is a good proxy to identify a patch\footnote{This patch is the one associated with the similar test case that failed in the past.} that will be more similar than the average patch in a dataset.

\begin{figure}[!t]
\centering
	\includegraphics[width=0.8\linewidth]{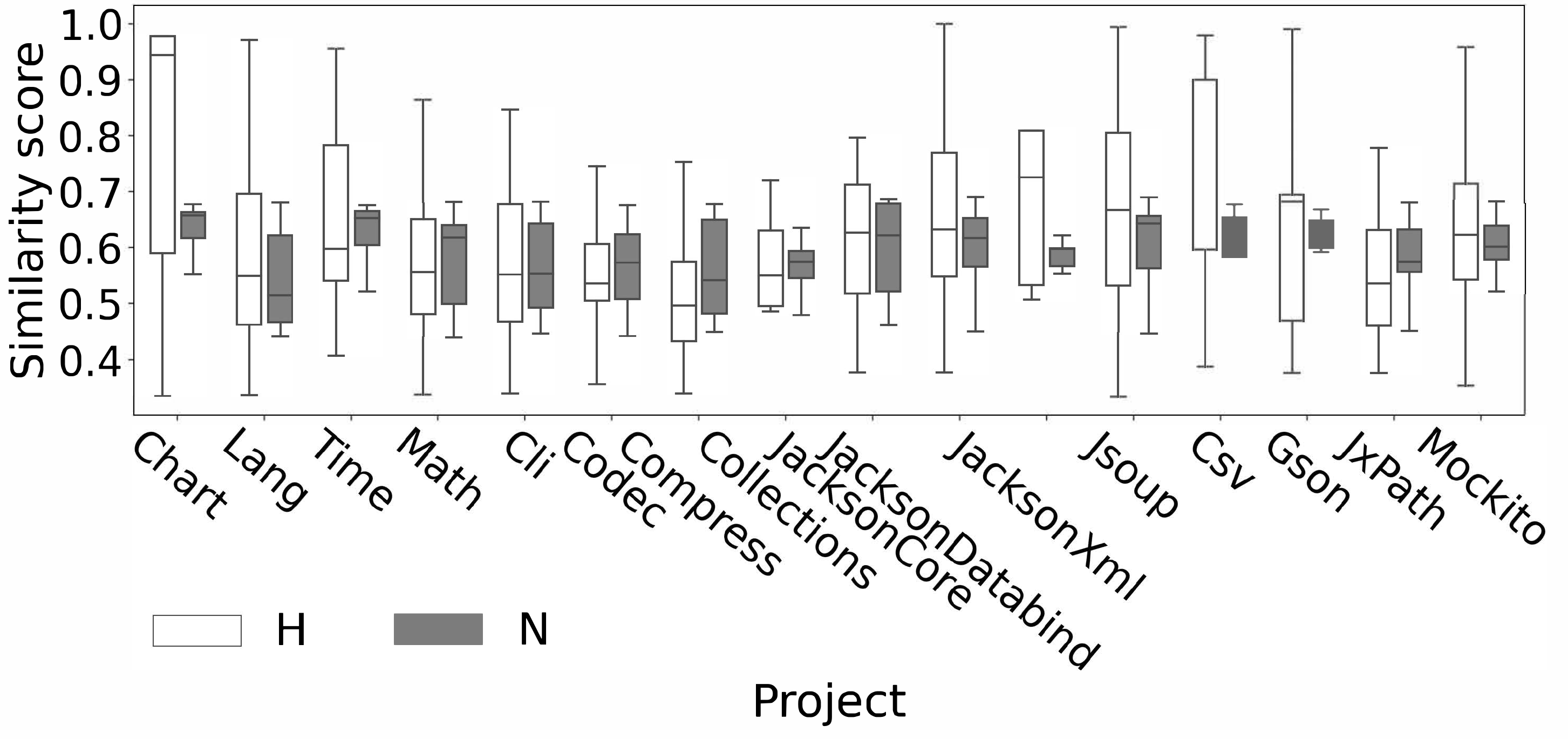}
	\caption{Distributions on the similarities of pairwise patches (similar patch selected with Scenario H vs. Scenario N from \underline{\bf other projects}, i.e., the search space for searching similar cases does not include the buggy project itself).}
	\label{fig:boxplot_nofilter2}
\end{figure} 

In the aforementioned experiment, the test cases in search space are allowed from the same project due to the lack of test cases. 
To evaluate our hypothesis in the scene of insufficient test cases, we further reproduce the comparison by focusing on test cases and patches that are from other projects (i.e., the buggy project itself is excluded from the search space of test cases). 
Figure~\ref{fig:boxplot_nofilter2} further provides the distributions for the two scenarios H and N. 
In this case, we note that the difference is less pronounced for most projects. We postulate that this is due to the fact that similarity scores are low. Thus we propose to set a threshold and consider, for the scenario H, cases where the test cases present a higher similarity than the threshold.

\begin{figure}[!t]
\centering
	\includegraphics[width=0.8\linewidth]{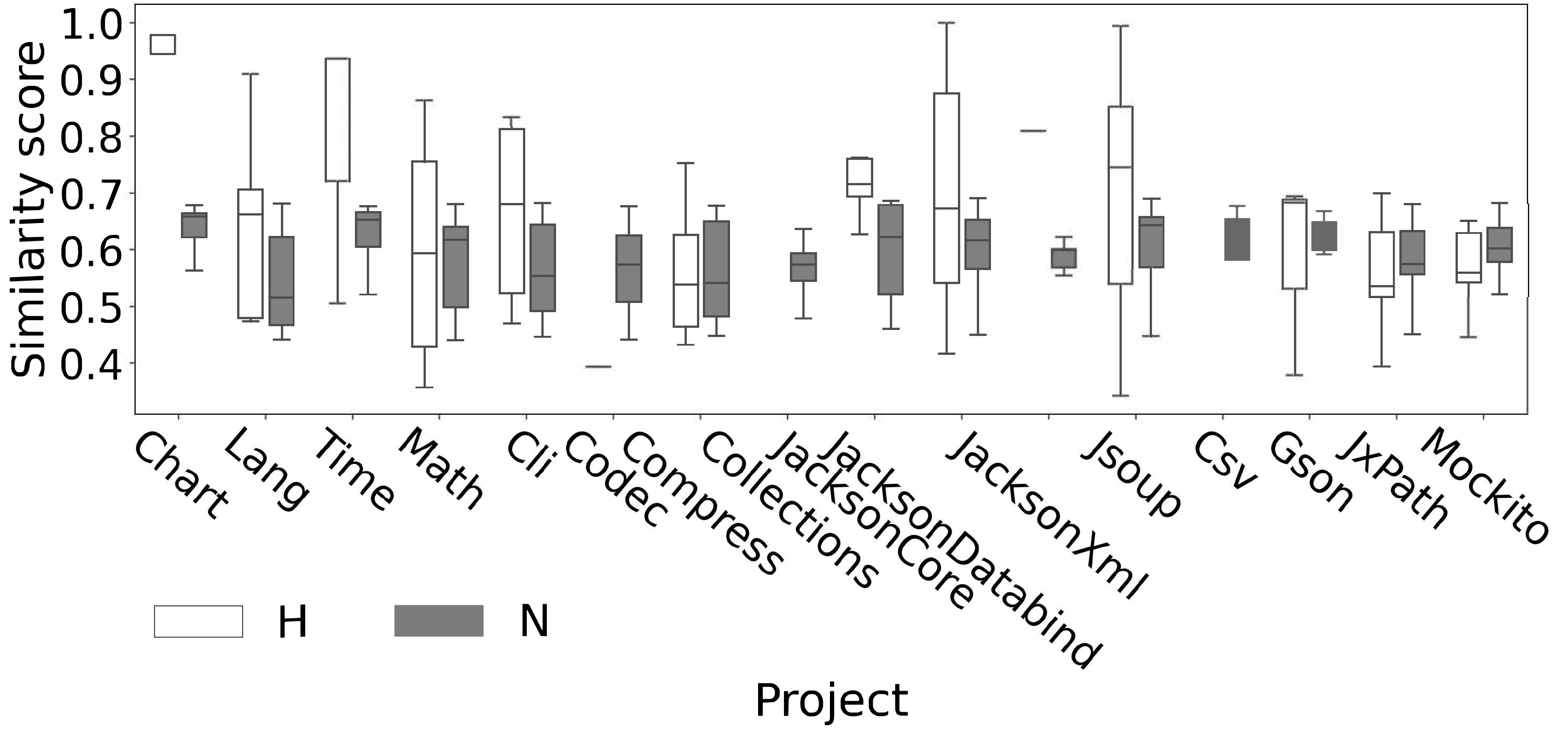}
	\caption{Distributions on the similarities of pairwise patches (similar patch selected with Scenario H vs. Scenario N from {\underline{\bf other projects}}, i.e., the search space for searching similar cases does not include the buggy project itself, by setting the threshold at 0.6).}
	\label{fig:boxplot_filter}
\end{figure}

When considering each of the 1,120 test cases, for many of them the closest test case in the search space is actually not a ``similar'' one. 
By looking at the pairwise similarity distribution shown in Figure~\ref{fig:Similarity_Test}, we note that 74\% (831/1120) similarity values are lower than 0.6. 
We therefore arbitrarily decided to use this value as the threshold\footnote{Note that it aims to validate our hypothesis but not to infer a specific/adaptable threshold.} to explore the hypothesis by isolating the minority of cases where the similarities are significant (more experiments with different thresholds are presented in Section~\ref{subsec:rq2} for RQ-2).
Figure~\ref{fig:boxplot_filter} presents the comparison of similarity distribution when the patch pairs are selected with Scenarios H and N from other projects, after setting a threshold of 0.6 for the similarity of test cases in Scenario H to reduce noise. 
We observe that scenario H \underline{now} provides the highest similarities for the paired patches (based on the similarities of test cases). Note that some white boxes (scenario H) in the plot are missing is due to lack of high enough similar test cases. %

\find{{\bf[RQ-1.2] } Given a test case and its most similar test case, their associated patch pair will exhibit a similarity that is statistically higher than the average similarity for all pairs of patches in the same project. This finding is also confirmed across projects.}

\subsection{[RQ-2] Identifying Correct Patches with \toolname}
\label{subsec:rq2}

\noindent
{\bf Objective:} 
Findings in answering the above RQ confirm our hypothesis that test case similarity correlates with patch similarity. \toolname is therefore implemented to explore this hypothesis scenario in an APR setting where generated plausible patches (for a failing test case $T$) are ranked based on their similarity with a set of historical patches that were applied by developers (to address failing test cases similar to $T$). To answer this RQ, we leverage the 1,278 plausible patches which are composed of 205 developer-written patches and 1,073 APR-generated patches by the tools in Table~\ref{tab:patches}.

\abox{By assessing \toolname on the collected dataset of plausible patches generated by literature APR tools, our main aim is to demonstrate to the community the feasibility of the proposed research direction for patch assessment.}

\noindent
{\bf [Overall Assessment]:} We first design a baseline with a simple hypothesis which considers that a patch is more likely to be correct if it is similar to some historically correct patches (i.e., any correct patches). In contrast to \toolname, this baseline does not consider failing test cases as the constraint for reducing the search space. We recall that the performance of \toolname, which relies on search, is dependent on the availability (in project repositories) of test cases that are actually similar to the failing test case addressed by the APR-generate patches. As introduced earlier, the closest test cases in the search space may actually not be that similar: this is a classical challenge of search engines~\cite{liu2021opportunities}. Therefore, we propose to filter in only test cases that present a sufficient level of similarity with the targeted test cases. Experimental evaluations will further offer insights on the use of such a threshold.

We chose the cosine similarity for the baseline and \toolname implementation.
Table~\ref{tab:AUC_baseline} reports the classification performance of the baseline (AUC less than 0.6). For reference, the performance of \toolname is provided later (cf. Table~\ref{tab:AUC_cosine}). 
We note that the baseline performance is similar with \toolname when the test similarity threshold is not set.
However, when we only consider test cases with higher similarity with the failing test cases, the performance of \toolname increases (up to 0.85 for +Recall and 0.71 AUC). CC2Vec, as an embedding model for patches, helps achieve high AUC, F1, +Recall (i.e., the recall in identifying correct patches) and -Recall (i.e., the recall in filtering out incorrect patches). 

\begin{table}[!h]
	\centering
	\caption{Baseline's performance on identifying (in)correct patches.}
	\label{tab:AUC_baseline}
	{
	\begin{threeparttable}
		\begin{tabular}{lrr|cccc}
			\toprule
		\makecell[l]{{\bf Patch}\\ {\bf embedding$^\dagger$}}
        & {\bf Correct} & {\bf Incorrect} 
		& {\bf AUC} & {\bf F1} & {\bf +Recall} & {\bf -Recall} \\
			\midrule
			 CC2Vec & \multirow{2}{*}{589} & \multirow{2}{*}{689} & 0.586 & 0.579 & 0.705 (415) & 0.379 (261) \\
			 Bert & & & 0.593 & 0.558 & 0.647 (381) & 0.428 (295) \\
			\bottomrule
		\end{tabular}
			
	\end{threeparttable}
	}
\end{table}

\begin{table}[!h]
	\centering
	\caption{\toolname's performance on identifying (in)correct patches.}
	\label{tab:AUC_cosine}
	{
	\begin{threeparttable}
		\begin{tabular}{lrrr|cccc}
			\toprule
		\makecell[l]{{\bf Patch}\\{\bf  embedding$^\dagger$}} 
		& \makecell[l]{{\bf T$^\ast$}}
        & \makecell[r]{{\bf \# Correct}\\{\bf  patches}} & \makecell[r]{{\bf \# Incorrect}\\{\bf  patches}} 
		& {\bf AUC} & {\bf F1} & {\bf +Recall} & {\bf -Recall} \\
			\midrule
			 \multirow{6}{*}{CC2Vec}& 0.0 & 589&689 & 0.557 & 0.549 & 0.628 (370) & 0.437 (301)  \\
			 &0.6 & 144&181 & 0.559 & 0.505 & 0.562 (81) & 0.470 (85)  \\
			 &0.7 & 94&141 & 0.678 & 0.590 & 0.766 (72) & 0.447 (63)  \\
			 &0.8 & 57&57 & \cellcolor{blue!25} 0.718 & \cellcolor{blue!25} 0.722 & \cellcolor{blue!25} 0.842 (48) & \cellcolor{blue!25}0.509 (29) \\
			 &0.9 & 41&44 & 0.709 & 0.693 & \cellcolor{black!25} 0.854 (35) & 0.432 (19)  \\
			\midrule
			 \multirow{6}{*}{BERT}& 0.0 & 589&689 & 0.561 & 0.518 & 0.593 (349) & 0.406 (280) \\
			 &0.6 & 144&181 & 0.611 & 0.576 & 0.694 (100) & 0.431 (78) \\
			 &0.7 & 94&141 & 0.639 & 0.570 & 0.766 (72) & 0.440 (62) \\
			 &0.8 & 57&57 & 0.676 & 0.626 & 0.719 (41) & 0.421 (24)\\
			 &0.9 & 41&44 & 0.647 & 0.600 & 0.732 (30) & 0.341 (15) \\
			\bottomrule
		\end{tabular}
		{\footnotesize{$^\dagger$}Embeddings of test cases are always done with code2vec. \\
		$^\ast$T: Threshold of test case similarity. Given the failing test case of an APR-generated patch, we consider only historical test cases with the similarity which are higher than the threshold. Thus, depending on the threshold, some generated patches cannot be assessed as we are not able to associate them with any past test case.\\
		``(\#)'' in last two columns represents the number correct/incorrect patches identified by \toolname.}
	\end{threeparttable}
	}
\end{table}

Table~\ref{tab:MAP_MRR} provides performance results in terms of MAP and MRR. The high metric values further confirm that most correct patches are indeed ranked higher in the recommended patch list that is sorted based on their similarities with historically-relevant patches (given test case similarity).
The MAP and MRR of the baseline are both 0.63, underperforming against \toolname (up to 0.80 and 0.8 for MAP and MRR respectively).
\begin{table}[!t]
	\centering
	\caption{\toolname's performance on ranking correct patches.}
	\label{tab:MAP_MRR}
	{
	\begin{threeparttable}
		\begin{tabular}{lrrrrr}
			\toprule
			{\bf Threshold of test case similarity} & {\bf 0.00} & {\bf 0.60} & {\bf 0.70} & {\bf 0.80} & {\bf 0.90} \\
			\midrule
			\bf MAP  & 0.62  & 0.63  & 0.71  & 0.80  & 0.70  \\
			\bf MRR & 0.63  & 0.65  & 0.74  & 0.81  & 0.75  \\
			\bottomrule
		\end{tabular}
	\end{threeparttable}
	}
\end{table}

\begin{figure}[!t]
\centering
	\includegraphics[width=0.65\linewidth,]{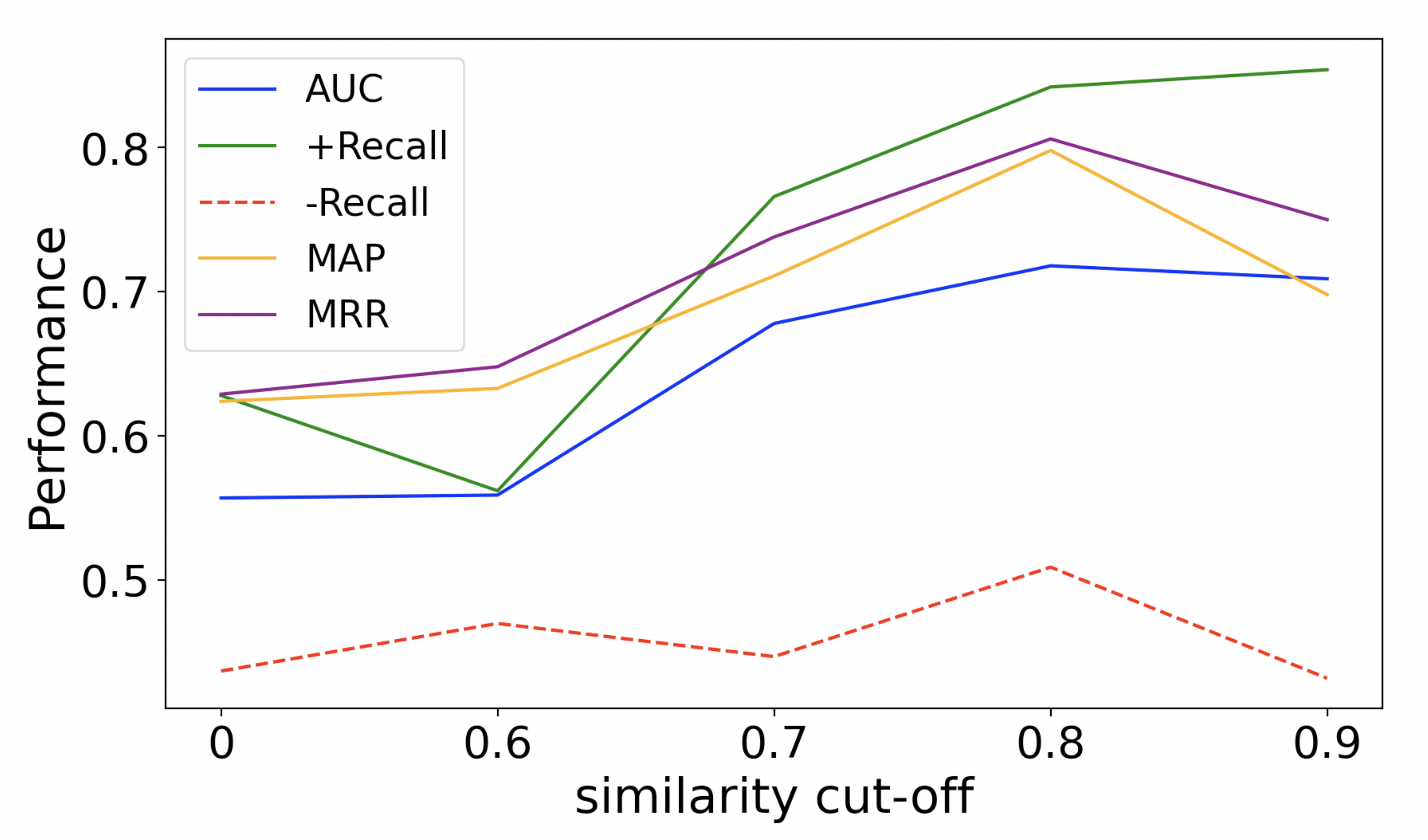}
	\caption{Performance evolution of \toolname with varying threshold of the test-case similarity.}
	\label{fig:change}
\end{figure}

Figure~\ref{fig:change} illustrates the overall performance evolution of \toolname when the threshold of the test case similarity is varied. We note, while the -Recall does not change drastically, there is a positive effect on +Recall (and other metrics) when the similarity threshold is increased.  These results confirm that the underlying hypothesis of \toolname is just: when \toolname identifies highly similar test cases, its prediction of correctness for APR-generated patches is more accurate.

{\bf Representation learning vs. raw strings}
Beyond the performance metrics on AUC, +/-Recall and F1 of \toolname, we propose to investigate the choice of representation learning for embedding patches. 
To assess the value of leveraging deep learning models for embedding patches, we consider computing similarity of patches as raw strings. To that end, we simply use the Levenshtein distance between the patches (considered as strings, and not after computing embeddings): by setting the threshold of test case similarity as 0.8, we obtain the lowest AUC and F1 values, respectively at 0.46 and 0.53. -Recall even drops at 0.12. These results validate our decision to leverage deep representation learning models for producing patch embeddings.

\find{{\bf[RQ-2]} \toolname performs well in identifying patch correctness with an AUC $\sim$ 0.7 and an F1$\sim$0.7 while the Recall of identifying correct patches reaches $\sim$0.8. Further experimental investigations, with constraints on test case similarities, support our hypothesis: similar test cases are addressed by similar patches.}

\subsection{[RQ-3] Competitive/Complementary to the State-of-the-art}
\label{subsec:rq3}

{\bf Objective:} In previous research questions, we validate our hypothesis and develop the pipeline \toolname to identify the correctness of patches generated by APR tools. The experimental results show \toolname achieves promising performance. To further evaluate \toolname in practice, we compare it against the state of the art static and dynamic approaches with the same dataset used for RQ-2.
Recall that, in practice, the performance of \toolname, is impacted by the availability to find (really) similar test cases. As illustrated in the results of Table~\ref{tab:AUC_cosine}, when the threshold of test case similarity is set high, the number of patches for which similar test cases are found in our dataset decrease significantly. While this does not contradict the hypothesis underlying \toolname, it may limit its value when larger datasets are not available. Nevertheless, we postulate that \toolname can be complementary to {the previous two} state of the art approaches (i.e., Tian et al.~\cite{tian2020evaluating} and PATCH-SIM~\cite{xiong2018identifying}).

\subsubsection{\bf Comparing Against the State-of-the-art} We conduct the comparison of \toolname against the state-of-the-art patch correctness predicting tools, i.e., Tian~{\em et al.}'s deep learning approach~\cite{tian2020evaluating}, PATCH-SIM~\cite{xiong2018identifying} and ODS~\cite{ye2021automated}.

\subsubsection*{\bf \toolname vs. Deep learning approach}
Since \toolname leverages pre-trained DL-based (deep learning representation) models, we proceed to compare it against the recent related work by Tian~{\em et~al.}~\cite{tian2020evaluating} where DL-based embeddings of patches are leveraged for static checks. To ensure that the results of \toolname and of the DL-based approach are compared on the same dataset, we focus on the 114 patches (threshold>0.8) as the testing set of the DL-based approach. Because the approach of  Tian~{\em et~al.} require a training dataset for supervisely producing a classifier, we use the remaining 1164 (1278-114) patches. As for classification algorithms, we leverage both Logistic Regression (LR) and Random Forrest (RF), following the experiments of the authors. As shown in in Table~\ref{tab:AUC_LR_RF}, \toolname can recall (+Recall) more correct patches while preserve the same or higher -Recall. 
It should be noted also that, unlike the DL-approach by Tian~{\em et al.}, \toolname doesn't require large dataset for training a classifier.
\begin{table}[!h]
	\centering
	\caption{Comparison with a state of the art supervised classifier~\cite{tian2020evaluating}.}
	\label{tab:AUC_LR_RF}
	{
	\begin{threeparttable}
		\begin{tabular}{lrrrr}
			\toprule
			{\bf Classifier} & {\bf AUC} & {\bf F1} & {\bf +Recall} & {\bf -Recall} \\
			\midrule
			\multirow{1}{*}{Tian et al. (LR)}   & \cellcolor{black!25}0.72 & 0.68 & 0.77 & \cellcolor{black!25}0.51  \\
			\multirow{1}{*}{Tian et al. (RF)}  & 0.70 & 0.49 & 0.75 & 0.46  \\
			\midrule
			\multirow{1}{*}{\toolname} & \cellcolor{black!25}0.72 & \cellcolor{black!25}0.72 & \cellcolor{black!25}0.84 & \cellcolor{black!25}0.51 \\
		   \bottomrule
		\end{tabular}

	\end{threeparttable}
	}
\end{table}

\subsubsection*{\bf \toolname vs. PATCH-SIM}
The state of the art in dynamic assessment of patch correctness is PATCH-SIM~\cite{xiong2018identifying}. It targets excluding incorrect patches via comparing execution behaviour of tests for the patched and original programs. We apply PATCH-SIM to the 114 test data to generate prediction. However, PATCH-SIM fails to produce prediction results for some of the bugs/patches\footnote{We reported the problem to the PATCH-SIM authors and we are still waiting for their response.}. Furthermore, we observed that PATCH-SIM takes several hours to produce a prediction outcome for some patches. In our experiment, we set a one hour timeout for the prediction per patch. Eventually, 48 out of 114 could be evaluated by PATCH-SIM. Our experiment was conducted on Linux server equipped with 8 cores 2.10GHz CPU and 125G memory. Results in Table~\ref{tab:patch-sim} show that PATCH-SIM achieves a 0.80 of +Recall and a 0.42 of -Recall. Among patches evaluated, the average prediction time for each patch is 1044s (i.e., more than 17 minutes), while \toolname only spends $\sim$0.3 second on validating each patch.
Overall, the dynamic approach, PATCH-SIM, is constrained in terms of resource requirements, as it needs to generate new test inputs and exploit the behavior similarity of test case executions for validating each patch.
The resource cost of \toolname is mainly decided by two aspects: \ding{182} the model training process, although \toolname leverages the pre-trained models, and \ding{183} the search space of historical test cases. The time cost of finding similar test cases will be sharply increased only when the search space is expanded.

\begin{table}[!h]
	\centering
	\caption{Comparison with a state of the art dynamic-based patch assessment~\cite{xiong2018identifying}}
	\label{tab:patch-sim}
	{
	\begin{threeparttable}
		\begin{tabular}{lrrrrr}
			\toprule
			{\bf Classifier} & {\bf AUC} & {\bf F1} & {\bf +Recall} & {\bf -Recall}  \\
			\midrule
			\multirow{1}{*}{PATCH-SIM} & 0.61  & 0.52  & 0.80 & 0.42 \\
			\multirow{1}{*}{\toolname} & \cellcolor{black!25}0.72  & \cellcolor{black!25}0.72  & \cellcolor{black!25}0.84 & \cellcolor{black!25}0.51 \\
		   \bottomrule
		\end{tabular}

	\end{threeparttable}
	}
\end{table}

\subsubsection*{\bf \toolname vs. ODS}
\label{subsubsec:ods}
We also compare our performance against a recent machine learning-based approach leveraging manually-engineered features. 
We compare \toolname against a recent work by Ye {\em et~al.} \cite{ye2021automated} where the authors propose a supervised learning approach ODS that explores manually engineering patch features for overfitting detection. To predict patch correctness, ODS first constructs the engineering features from the AST representation of patches to express potential behavior information. Such features are used by ODS to train a ML-based model to proceed the classification of patch correctness.
While we could not fully reproduce their work on our dataset due to the unavailability of their training dataset, we are able to compare our results with the ones presented in their paper since we have test sets of similar size and from the same sources. Overall, \toolname and ODS exhibit similar performance metrics. When they tune their learners to have a high +Recall (e.g., 1.00), their -Recall drops (e.g., 0.46, respectively). Our \toolname unsupervised learning approach further aims to cope with two challenges with approaches such as ODS: (1) they require large sets of labelled patches to perform supervised learning;  (2) it can be difficult to manually explain a prediction classification (e.g., because features that contribute to the classification decision are difficult to track back to the failing test case specification and may not generalize to new data). 

\find{{\bf[RQ-3]} \ding{182} When the availability of similar test cases is satisfied, \toolname can achieve competitive performance on predicting the correctness of APR-generated patches against the state-of-the-art dynamic and static approaches.}

\subsubsection{\bf Enhancing the State-of-the-art with \toolname}

{We present experimental results to demonstrate that we can achieve enhanced performance in correct patch identification by using existing (static or dynamic) state of the art approaches in conjunction with \toolname.}

\vspace{2mm}
\subsubsection*{\bf\em I. Supplementing a supervised classifier with \toolname}~\newline
\label{sec:bats-ml}
\noindent
{\bf Objective:} Given that \toolname is fairly accurate when highly similar test cases (with associated historical patches) are available, we propose to build a pipeline where \toolname is applied on the subset of bug cases where such test cases exist. For the rest of bugs, the correctness of generated patches is predicted by using a relevant literature classification-based approach proposed by Tian~{\em et~al.}~\cite{tian2020evaluating}. We then compare the performance of this pipeline against the performance yielded when the classifier is used alone on the whole dataset. 

\noindent
{\bf Experiments:} We set a threshold of the test case similarity at 0.6 (cf. Table~\ref{tab:AUC_cosine}) to identify which failing test cases are relevant for assessing the added-value of \toolname. The dataset is then split into 325 patches for test and 935 for training a patch classifier described in recent literature:  we reproduce the work of Tian {\em et~al.}~\cite{tian2020evaluating} using their provided artefacts which provide a supervised learning model to classify patches based only on embeddings (computed with Bert). As for learner, we leverage alternatively Logistic Regression (LR) and Random Forrest (RF) following the experiments of the authors. 

We first compute the performance of the supervised learning classifier alone on the test dataset. Then, we evaluate the combined pipeline (Tian {\em et~al.}~\cite{tian2020evaluating} + \toolname) with the following procedure: \toolname is first applied to predict correctness for all patches that are associated to test cases for which historical test cases with a high similarity ($\geq0.9$) can be found. 26.1\% (85/325) of patches in the test set are then evaluated by \toolname. The rest of patches are passed to the trained supervised classifier which does not require test case information.

\noindent
{\bf Results:} The results presented in Table~\ref{tab:AUC_LR_RF_combine} show that the combined pipeline can indeed supplement the baseline classifier. +Recall can be improved by up to 5 percentage points while -Recall can be improved by up to 4 percentage points.

\begin{table}[!t]
	\centering
	\caption{Supplementing a supervised classifier with \toolname.}
	\label{tab:AUC_LR_RF_combine}
	{
	\begin{threeparttable}
		\begin{tabular}{lrrrr}
			\toprule
			{\bf Classifier} & {\bf AUC} & {\bf F1} & {\bf +Recall} & {\bf -Recall} \\
			\midrule
			\multirow{1}{*}{Tian et al. (LR)} & 0.75 & 0.60 & 0.53 & 0.80  \\
		    \multirow{1}{*}{Tian et al. (LR) + \toolname} & 0.75 &   \cellcolor{black!25} 0.62 & 0.53 & \cellcolor{black!25}0.85  \\
            \midrule
           \multirow{1}{*}{Tian et al. (RF)} & 0.75  & 0.59 & 0.56 & 0.82  \\
           \multirow{1}{*}{Tian et al. (RF) + \toolname} & 0.75 &  \cellcolor{black!25}0.67 &  \cellcolor{black!25}0.61 & 0.82  \\
         
		   \bottomrule
		\end{tabular}
	\end{threeparttable}
	}
\end{table}

\find{{\bf[RQ-3]} \ding{183} \toolname can supplement a state of the art patch classification system, which : on bug cases where the failing test case is highly similar to historical test cases, \toolname can provide more accurate classification.}

\vspace{2mm}
\subsubsection*{\bf\em II.  Complementing test execution based detection}~\newline
{\bf Objective:}  Our objective is to assess whether \toolname (which statically reasons about test case similarity) can boost PATCH-SIM~\cite{xiong2018identifying} (which considers dynamic execution behaviour). We consider that \toolname value can be confirmed if it can help exclude incorrect patches that PATCH-SIM could not. Therefore, we build on a similar pipeline than in Section~\ref{sec:bats-ml}, where \toolname is applied instead of PATCH-SIM when enough similar test cases can be found.

\noindent
{\bf Experiments:}
We apply PATCH-SIM to the above 325 test patches in last Section~\ref{sec:bats-ml}. As for \toolname, we use the Defects4J dataset as search space. Note that, as in all experiments, we ensure that the search space does not include test case or patches linked to the assessed generated patch. To achieve reliable performance with \toolname, we must consider only cases where highly similar test cases exist in our dataset. Therefore it is possible that our performance measurement could only be computed on a portion of the test set. Finally, PATCH-SIM can successfully produce results for 153 patches, of which 48 patches are applied by \toolname when setting the similarity threshold at 0.8. is sufficient to find similar test cases. 

\noindent
{\bf Results:}
Table~\ref{tab:patchsim_bats_improvement} presents the performance results of PATCH-SIM (alone) and the combination (PATCH-SIM with \toolname). We note that, by applying \toolname to the subset of patches where similar test cases are available, we are able to improve the overall performance in patch correctness prediction. Theses results  demonstrate that \toolname can be complementary to a dynamic approach. 

\begin{table}[!h]
	\centering
	\caption{Supplementing PATCH-SIM with \toolname.}
	\label{tab:patchsim_bats_improvement}
	{
	\begin{threeparttable}
		\begin{tabular}{lrrrr}
			\toprule
			{\bf Classifier} & {\bf AUC} & {\bf F1} & {\bf +Recall} & {\bf -Recall} \\
			\midrule
			\multirow{1}{*}{PATCH-SIM}  & 0.62 & 0.55 & 0.82 & 0.43 \\
		    \multirow{1}{*}{PATCH-SIM + \toolname} & \cellcolor{black!25}0.65 & \cellcolor{black!25}0.59 & \cellcolor{black!25}0.84 & \cellcolor{black!25}0.51  \\
		   \bottomrule
		\end{tabular}

	\end{threeparttable}
	}
\end{table}

\find{{\bf [RQ-3]} \ding{184} \toolname static approach is complementary to the PATCH-SIM~\cite{xiong2018identifying} dynamic approach. By being able to identify incorrect and correct patches when test cases are sufficient, \toolname implementation confirms our initial hypothesis that statically reasoning about similar test cases offers a novel and promising perspective to the assessment of APR-generate patch correctness.}

\section{Ablation Study}

\subsection{Bug types of failing test cases clusters} 
In this study, we manually check whether bugs, grouped with respect to the test cases, in each cluster are actually similar or not in terms of bug types. We first note that the failing test cases grouped in the same cluster perform similar checks (e.g., date-related checks). Building up on the dissection study of Defects4J bugs by Sobreira{\em et~al.}~\cite{sobreira2018dissection}, we further investigate the categories of bugs in each cluster. We find that test cases in a given cluster are indeed related to bugs in the same category. For instance, test cases triggering Chart-2 and Chart-4 bugs are in the same cluster and the dissection study data indicates that these two bugs  are in the same category of \texttt{NullPointerException}. However, it is noteworthy that the categories in the dissection proposed by Sobreira{\em et~al.} are too coarse-grained. Several of our clusters therefore relate to the same category. Finally, note that the dissection study was performed in a previous (smaller) version of Defects4J, which does not allow us to provide comprehensive results for our dataset. Nevertheless, the observations that we have made support the framing of our hypothesis that test cases similarity can reflect very well the category of bugs, and hence of the required fixes. In future work, we will consider exploring other possible representations of bugs beyond test cases, such as bug reports.

\subsection{Asymmetry of the hypothesis}
The experimental results validate the feasibility of our proposed hypothesis: similar test cases can lead to similar patches for fixing associated bugs.
We investigate the symmetry of the hypothesis: {\em Could similar patches be referred to similar test cases}? 
To this end, we first independently cluster patches with their similarities, and then assess the similarities of the associated test cases in each patch cluster.
With respect to independent clustering of patches, the clustered patches achieve a Cluster Similarity Coefficient ($CSC$) of 0.26 and all the groups are qualified. However, the $CSC$ for associated test cases groups is 0.03 that is very close to zero and much lower than the $CSC$ value of clustered patches.
The results indicate that the symmetry of the hypothesis is invalidated, i.e., similar patches cannot fully be mapped to addressing similar failing test cases.
It is reasonable, since different bugs can be fixed with similar code change ways which lead to similar patches~\cite{liu2021mining}, but the different bugs are triggered by different test cases.
\section{Threats to Validity}

\noindent
{\sc \em Threats to External Validity.} 
We relied on Bisecting-K-means for the clustering experiments to validate our hypothesis. Other algorithms may reveal different results. We have mitigated a potential bias by using multiple evaluation metrics to exhaustively assess the clusters. We also relied only on Defects4j to ensure that we can collect enough plausible patches from literature APR experiments. 

In future work, the community could further investigate other datasets  as well as test cases augmentation (through test generation~\cite{peng2020automated,soremekun2021probabilistic,selakovic2018test} or code search~\cite{kim2018facoy}) to enlarge the datasets of patches and test cases.

\vspace{3mm}
\noindent
{\sc \em Threats to Internal Validity}
A major threat to internal validity is that we manually process patches to build the dataset. We may have introduced some mismatching errors when associating test cases. To mitigate this threat, we publicly release all artefacts for review by the community.

Towards reasoning about code similarity, we rely on code2vec, which parses test cases to deep learning features. Unfortunately, while most projects format their test case specifications as for typical code (such as in Figure~\ref{fig:Chart-26}), the Closure project presents an ad-hoc format where the essential parts of the specification are formatted as a string (see Figure~\ref{fig:closure-49}): in such cases, code2vec embeddings abstract away the string as a mere argument to a function, rendering the embeddings semantically irrelevant.
Therefore, because we leverage off-the-shelf tools to validate our hypothesis, we simply discard Closure bugs, for which future work can investigate specific learners to parse test specification defined in the form of string. Eventually, our experimental evaluation considers 1,278 plausible patches, 598 of which are correct while 689 are overfitting (i.e., incorrect). Overall, the dataset is, to the best of knowledge, the largest set of plausible patches explored in the literature on patch correctness assessment.

\begin{figure}[!t]
    \centering
\begin{lstlisting}[language=Java,frame=tb]
public void testDrawWithNullInfo() {
   boolean success = false;
      try {
         BufferedImage image = new BufferedImage(200, 100, 
         BufferedImage.TYPE_INT_RGB);
         Graphics2D g2 = image.createGraphics();
         this.chart.draw(g2, new Rectangle2D.Double(0, 0, 200, 100), null, null); g2.dispose();
      }
   catch (Exception e) {
         success = false;
   }
   assertTrue(success);
}
\end{lstlisting}
    \caption{A typical failing test case specification (Chart-26).}
    \label{fig:Chart-26}
\end{figure}

\begin{figure}[!t]
    \centering
    \scriptsize
\begin{lstlisting}[language=Java,frame=tb]
public void testComplexInlineNoResultNoParamCall3() {
       test("function f(){a();b();var z=1+1}function _foo(){f()}",
              "function _foo(){{a();b();var z$$inline_0=1+1}}"); 
\end{lstlisting}
    \caption{String-based format for test specification ({Closure-49}).}
    \label{fig:closure-49}
\end{figure}

\vspace{3mm}
\noindent
{\sc \em Threats to Construct Validity.}
The used embedding models are pre-trained and some parameter weights may not be adapted to our work. In future work, we could retrain and fine-tune the parameters after collecting large datasets. %

\section{Related Work}
\label{sec:relatedWork}

\paragraph{\bf Automated Program Repair.}
Automated program repair (APR) aims to liberate program developers from the manual debugging burdens by generating patches automatically, and has attracted significant attention as well as achieved promising performance on fixing bugs~\cite{goues2019automated,monperrus2018automatic,liu2021critical}.
Generate-and-validate program repair is one of the commonly studied APR approaches that generates a set of candidate patches and validates the potential patch by checking the patched programs with static analysis or test suites~\cite{long2016analysis}.
 Such approaches studied in the literature can be generally summarized into three categories: \ding{172} Heuristic based approaches, that construct and iterate over a search space of syntactic program modifications to find the adequate patch for the given in a heuristic way~\cite{weimer2009automatically, le2012genprog, liu2019tbar, yuan2020arja, jiang2018shaping,wen2018context,liu2019avatar,saha2017elixir,ghanbari2019practical};
\ding{173} Constrain based approaches, building on constraint solving to synthesize transformations for patch generation~\cite{nguyen2013semfix,mechtaev2016angelix,xuan2016nopol,xiong2017precise};
and \ding{174} Learning based approaches which explore machine learning techniques to boosting program repair by learning correct code or natural code transformation~\cite{liu2018lsrepair,li2020dlfix,lutellier2020coconut,jiang2021cure,gupta2017deepfix,white2019sorting}.
These state-of-the-art APR works are challenged by their generated plausible~\cite{qi2015analysis} but incorrect patches that are just overfitting the test suites but not really fix the program bugs.
This work aims to boost the APR by proposing a new approach that is dedicated to validating the patch correctness for APR tools.

\paragraph{\bf Patch Correctness.}
The state-of-the-art APR techniques mainly rely on static analysis techniques or test suite validate to the assess the feasibility of patches generated by themselves, which however could lead to overfitting plausible patches that actually do not fix the bugs or even make the patched programs worse~\cite{qi2015analysis,smith2015cure}.
To address this challenge, practitioners have been studying efficient approaches to automate the identification of correct patches. 
Xiong {\em et al.}~\cite{xiong2018identifying} explored similarity of test case executions to heuristically assess the patch correctness, and implemented a patch validation technique, PATCH-SIM, based on a reasonable observation that a correct patch does not tend to change the behaviour of originally passing test cases but will revise the behaviour of originally failing test cases. 
Gao {\em et al.}~\cite{gao2019crash} propose to optimize the search space of patch candidates for APR tools by using the crash-freedom as the oracle to discard patch candidates which crash on the new tests. Their results show that the overfiting problem of patch validation in APR could be alleviated.
Yang {\em et al.}~\cite{yang2020exploring} also explored the difference between the runtime behaviour of developer’s and APR-generated patches. 
While the majority of automatic program repair approaches focus on dynamic execution to decide on the correctness of the patches, Csuvik {\em et al.}~\cite{csuvik2020utilizing} have shown the potential for machine learning techniques to be used in the identification of correct patches: by exploiting embedding models such as Doc2vec and Bert on code, they have found that patches that introduce new variables or make many changes are more likely to be incorrect. Ye {\em et al.}~\cite{ye2021automated} proposed to leverage manually engineered features to predict overfitting via supervised learning. They constructed an overfitting detection system ODS that trains classifiers for patch prediction by extracting code features from AST representation of buggy code and generated patch code. On the other hand, to assess the reliability of patch correctness prediction techniques, Le {\em et al.}~\cite{le2019reliability} constructed a gold set of correctness labels of APR-generated patches where several different author and automated patch assessment approaches are evaluated.

\paragraph{\bf Representation Learning.} Embedding is a key challenge for reasoning about similarity. Initial works on software engineering artefacts have explored models~\cite{devlin2019bert, le2014distributed} trained on natural language text (such as Wikipedia). BERT is a widely used bidirectional encoder representations model in textual tasks where it outperforms the state of the art by learning to obtain the conditional parameters. Recent works have however proposed specialized architectures to be trained on code. For example, Alon~{\em et~al.}~\cite{alon2019code2vec} proposed code2vec to capture the semantic properties of code snippets by learning distributed representation. Zhangyin {\em et~al.} presented a pre-trained CodeBert~\cite{feng2020codebert} that leveraged programming language and natural language data on code tasks. To learn the representation of code changes, Hoang~{\em et~al.}~\cite{hoang2020cc2vec} proposed an attention-based neural network model CC2Vec with the hierarchical structure to predict the difference between the removed and added code of the patch.

\paragraph{\bf Code Similarity and Code Clone Detection.}
Finding code that implements similar functionality is what \toolname does to find similar test cases and similar patches.
DeepSim~\cite{zhao2018deepsim} leverages deep learning on semantic features matrix representing control- and data-flow information to predict whether a given pair of functions implements similar functionality. 
Fang et al.~\cite{fang2020functional} introduced a new granularity level of the call-callee relationship to capture similar functionality while leveraging word embeddings and graph embeddings to train a deep neural network for the same task.

To detect code clones across different programming languages, CLCDSA~\cite{nafi2019clcdsa} extracts 9 syntactic source code features from the ASTs of different pieces of code and uses either Cosine similarity or trained neural network to detect clones.
\toolname also uses Euclidean distance and Cosine similarity to find similar test cases and similar patches.
Alternatively, SLACC~\cite{mathew2020slacc} uses dynamic analysis and input-output pairs to detect clones across different programming languages.
Finally, binary code similarity has been studied extensively~\cite{haq2021survey} with recent approaches also leveraging graph and instruction embeddings~\cite{feng2016scalable,xu2017neural,gao2018vulseeker} to find similar binaries.
These approaches for finding similar code across different programming languages and across binaries could benefit \toolname in the future by enabling cross-project and cross-language search for similar test cases and similar patches.

\section{Conclusion}
\label{sec:conc}
In this work, we propose and investigate a simple yet effective hypothesis for static patch correctness assessment: given a failing test case, any associated generated patch is likely correct if it is similar to patches that were used to address similar failing test cases. We have validated our hypothesis using the developer correct patches and the associated failing test cases in the Defects4J benchmark. To evaluate the potential of this hypothesis in predicting patch correctness, we propose a patch identification system, \toolname, to check patch behaviour against test specification based on unsupervised learning. \toolname achieves its highest performance when the similarity of test cases is high, which further validates our hypothesis. Compared against state of the art, \toolname outperforms them in identifying correct patches and filtering out incorrect patches. Despite potential issues in the availability of large datasets of projects to search for historical examples (test cases and their associated patches), we demonstrate that \toolname can be complementary to both state of the art static and dynamic approaches to predict patch correctness. Overall, \toolname offers insights on a promising research direction for patch assessment.

\section*{Acknowledgements}{
This work was supported by the funding from the European Research Council (ERC) under the European Union's Horizon 2020 research and innovation programme (grant agreement No. 949014), the National Natural Science Foundation of China (Grant No. 62172214), the Natural Science Foundation of Jiangsu Province (Grant No. BK20210279), and the Open Project Program of the State Key Laboratory of Mathematical Engineering and Advanced Computing (No. 2020A06).
}

\bibliographystyle{ACM-Reference-Format}%
\bibliography{sample-base}

\end{document}